\documentclass[aps,twocolumn,showpacs,floatfix,notitlepage]{revtex4-1}
\usepackage{amsmath,amsfonts,amssymb,graphicx,color,times,bbm,psfrag,dsfont,slashed,bm,hyperref}
\usepackage{slashbox}
\usepackage[T1]{fontenc}
\usepackage[latin9]{inputenc}

\usepackage[normalem]{ulem}
\newcommand{\bra}[1] {\langle{#1}\vert}
\newcommand{\ket}[1] {\vert{#1}\rangle}
\usepackage{tikz}
\newcommand{\ac}[1]{\textcolor{black}{#1}}

\newcommand\halfboxup[1]{
  \tikz[baseline=(n.base)]{\node(n)[inner sep=1pt]{$#1$};
    \draw[line cap=round](n.south west)--(n.north west)--(n.north east);
  }
  }

\begin{document}

\title{
Emerging 2D Gauge theories in Rydberg configurable arrays 
}

\author{Alessio Celi$^{1,2,3}$, Beno\^it Vermersch$^{1,2}$, Oscar Viyuela$^{4,5}$, Hannes Pichler$^{5,6}$, Mikhail D. Lukin$^{5}$, Peter Zoller$^{1,2}$}
\affiliation{1. Center for Quantum Physics, Faculty of Mathematics, Computer Science and Physics, University of Innsbruck, Innsbruck A-6020, Austria}
\affiliation{2. Institute for Quantum Optics and Quantum Information, Austrian Academy of Sciences, Innsbruck A-6020, Austria}
\affiliation{3. Departament de F\'isica, Universitat Autònoma de Barcelona, 08193 Bellaterra, Spain}
\affiliation{4. Department of Physics, Massachusetts Institute of Technology, Cambridge, Massachusetts 02139, USA}
\affiliation{5. Department of Physics, Harvard University, Cambridge, Massachusetts 02318, USA}
\affiliation{6. ITAMP, Harvard-Smithsonian Center for Astrophysics, Cambridge, Massachusetts 02138, USA}

\date{\today}

\begin{abstract}
{ 
Solving strongly coupled gauge theories in two or three spatial dimensions is of fundamental importance in several areas of physics ranging from high-energy physics to condensed matter. 
On a lattice, gauge invariance and gauge invariant (plaquette) interactions involve (at least) four-body interactions that are challenging to realize.  
Here we show that Rydberg atoms in configurable arrays realized in current tweezer experiments are the natural platform 
to realize scalable simulators of the Rokhsar-Kivelson Hamiltonian --a 2D U(1) lattice gauge theory that describes quantum dimer and spin-ice dynamics. 
Using an electromagnetic duality, we implement the plaquette interactions as Rabi oscillations subject to Rydberg blockade.
Remarkably, we show that by controlling the atom arrangement in the array we can engineer anisotropic interactions and generalized blockade conditions 
for spins built of atom pairs.    
 We describe how to prepare the resonating valence bond and the crystal phases of the Rokhsar-Kivelson Hamiltonian adiabatically, 
 and probe them and their quench dynamics by on-site measurements of their quantum correlations.  
 We discuss the potential applications of our Rydberg simulator to lattice gauge theory and 
 exotic spin models.   
}
\end{abstract}

\maketitle

\section{Introduction}

Atoms trapped in tweezer arrays and interacting via van der Waals
interactions of laser excited Rydberg states have recently emerged
as one of the most promising platforms for quantum simulation of spin
models. Unique features of Rydberg tweezer arrays include the flexibility of freely arranging atoms
in any geometric structure in one, two or three spatial dimensions \cite{Endres16,Barredo16,Barredo18}.
In combination with strong, and potentially angular-dependent, Rydberg-Rydberg
interactions, such arrays yield a versatile tool available to realize a wide
variety of effective spin models, as demonstrated in recent experiments
with Ising type \cite{Labuhn16,Bernier17,Lienhard18} and topological Su-Schrieffer-Heeger models \cite{Lienhard18ssh}
(for alternative realizations in optical lattices see \cite{Schauss15,Guardado18}). 
A key element of quantum many-body systems in Rydberg tweezer arrays is the Rydberg
blockade mechanism \cite{Browaeys16}. Here only single atomic Rydberg excitations within
a given blockade radius $R_{c}$ are allowed, with double excitations
strongly suppressed by large energy shifts from Rydberg van der Waals
interactions. In this paper, we show that such experimental setting provides
a natural framework for implementing 2D $U(1)$ lattice gauge models
for spin 1/2, and in particular (a variant of) 
the Rokhsar-Kivelson Hamiltonian \cite{Rokhsar88}.
Such Hamiltonian corresponds to a paradigmatic model of quantum spin ice and quantum dimers \cite{Moessner11}. 
Configurable 2D atomic
tweezer arrays thus offer a unique opportunity to study Rokhsar-Kivelson
dynamics and phase diagrams, in particular accessing and characterizing
its resonating valence bond phase \ac{\cite{Pauling49,Anderson73,Fazekas74}}.

The implementation of lattice gauge theories in spatial dimension larger than one is presently
one of the key challenges in the ongoing development of quantum simulators.
Recently, pioneering experiments have demonstrated quantum simulation
of 1D lattice gauge theories, including the 1D Schwinger model, as 1D quantum electrodynamics,
with trapped ions \cite{Martinez16,Kokail19}, and superconducting
qubits \cite{Klco18}. Furthermore, recent experiments with 1D Rydberg
chains \cite{Bernier17} could be reinterpreted in terms of a truncated 1D
Schwinger model \cite{Surace19} and string breaking. In contrast,
lattice gauge theories in higher spatial dimensions are much harder to simulate. 
They are expected to display a plethora of novel physical
phenomena which are absent in 1D, due to the interplay between electric
and magnetic interactions, such as confined-deconfined phase transitions
\cite{Polyakov87} and topological order \cite{Wen90}. A difficulty
in implementing lattice gauge theories in higher spatial dimensions is that gauge invariance (Gauss law) 
and gauge invariant magnetic interactions, \emph{plaquette} terms, typically
translate into 4-body (or higher order) interactions. This difficulty also applies
when gauge field excitations are represented as finite dimensional, such
as in lattice gauge spin models \cite{Horn81,Orland90,Chandrasekharan97,Moessner01b}. 
While recent proposals report significant advances in constructing gauge
invariant terms in Kogut-Susskind \cite{Kogut75} like Hamiltonians  from basic and natural
building blocks, e.g. in cold atom systems \cite{Zohar12,Banerjee12,Glaetzle14,Kasper17,Dutta17,Caldarelli17,Zache18},
a laboratory implementation of 2D lattice gauge theories remains elusive
(for a digital approach see e.g. \cite{Tagliacozzo13,Tagliacozzo13na,Zohar17,Bender18}, 
for reviews see \cite{Wiese13,Zohar15,Dalmonte16}, 
for Floquet engineering and related progress with density-dependent gauge fields see \cite{Barbiero18} and \cite{Greschner14,Clark18,Gorg18,Schweizer19}). 
For instance, plaquette (ring-exchange) interactions has been experimentally demonstrated 
for disconnected plaquettes only \cite{Dai17}.  

In this work we take a different route for achieving a natural implementation of 2D U(1) spin-1/2
models. The enabling insight is the existence of a \emph{dual formulation} where plaquette interactions
are mapped into single-body terms with constraints. In the context of our
Rydberg tweezer array these correspond to Rabi couplings between atomic
states which are subjected to generalized blockade conditions due
to Rydberg interactions. Thus, we obtain a natural relation between
the gauge theories and atomic systems with generalized blockade constraints,
which provides a physical basis for scalable \ac{\cite{notescalable}
} 
quantum simulation of
lattice gauge theories in 2D.

The idea of exploiting dualities for quantum simulation of spin models and lattice gauge theories is not new, see e.g. \cite{Grass16}. 
It is well known since \cite{Kogut75} that in the magnetic basis pure gauge Kogut-Susskind Hamiltonian simplifies 
(it is the same duality that relates $Z_2$ gauge theory like the toric code to the Ising model \cite{Wegner71,Fradkin78}) and 
allow, e.g., to rewrite Higgs-U(1) Hamiltonian \cite{Fradkin79} as an extended Bose-Hubbard model \cite{Kasamatsu13} (see also \cite{Bazavov15,Cuadra17}).
However, the duality we construct here for the pure gauge U(1) spin-1/2 models (quantum spin ice in the condensed matter language) has not been exploited for quantum simulation previously.  

The paper is organized as follows. In Sec. \ref{sec:dual}, we introduce the Rokhsar-Kivelson Hamiltonian as a 2D U(1) spin-1/2 gauge theory
and derive the dual Rokhsar-Kivelson Hamiltonian. First, after a brief tutorial on lattice gauge theories, 
we review the phase diagram of RK Hamiltonian on the square lattice without background charges as known in its original basis (Sec. \ref{sec:GM}).   
Then, we define the duality transformation, describe its properties, and illustrate the phase diagram in terms of the observables of the 
dual spins on the full square lattice and on ladder geometries (Sec. \ref{sec:duality}). 
In Sec. \ref{sec:imple}, we show how such dynamics can be naturally realized in 2D Rydberg arrays. 
First, we engineer 2D Ising models with tunable anisotropic interactions in \emph{decorated arrays}, obtained by arranging orientable pairs of Rydberg atoms on 2D arrays (Sec. \ref{sec:tunable_Ising}). 
The gauge theory emerges in such models for properly chosen arrangements (interactions) in the limit of small Rabi coupling (= transverse field), i.e. in a generalized blockade regime 
(Sec. \ref{sec:periodic_ladder}-\ref{sec:2Dgenblock}). 
We show that such Rydberg Rokhsar-Kivelson Hamiltonians host resonating valence bond phases and we propose a step-by-step prescription to prepare and detect these phases in current experiments (Sec.\ref{sec:prepa}). 
Finally in Sec. \ref{sec:outlook} we summarize our results and discuss future steps and potential applications of our simulator based on controllable Rydberg arrays to gauge theories and beyond.

\begin{figure}[ht!]
\begin{center}
\resizebox{0.99\columnwidth}{!}{\includegraphics{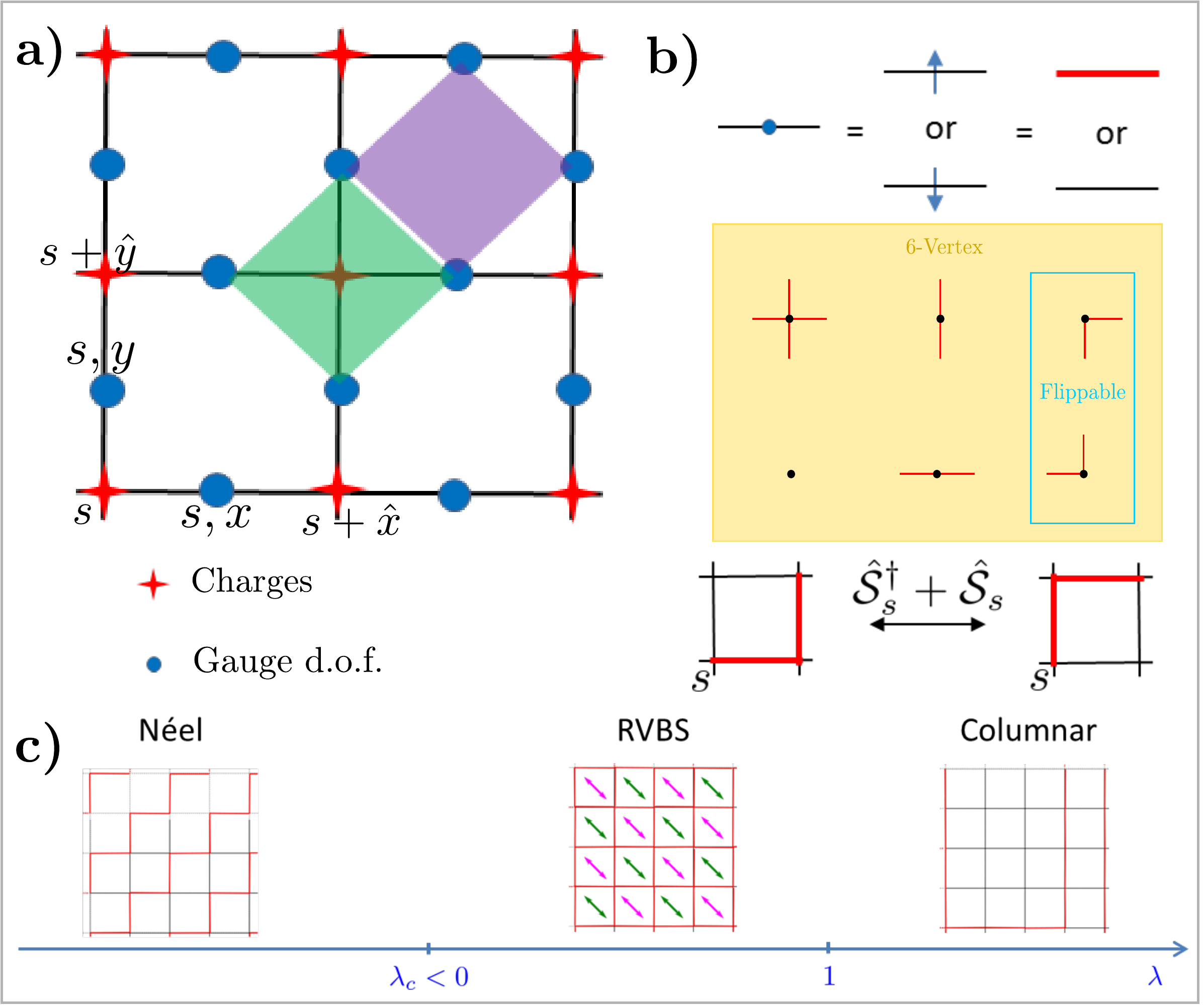}}
\caption{(color online) {\bf Spin gauge theories and the Rokhsar-Kivelson model.} 
{\bf a)}  Spin-1/2 lattice gauge theories are the simplest gauge theories with a continuous gauge symmetry group. 
The gauge d.o.f are spins 1/2 that live on the oriented links of the lattice, labeled as $s,\mu$, where $s$ is the starting site and $\mu$ is the direction, e.g. $x$ or $y$ 
in a square lattice.
They display the main features of gauge theories in $D>1$: i) the Gauss law that determines the allowed spin configurations on the links in terms of the
charges,  and ii) the magnetic interactions acting on the links around a plaquette. On a square lattice these operators correspond both to four-body spin operators indicated by the green and violet rhombi, respectively.      
{\bf b)} In a system without charges, we depict the effect of the Gauss law by coloring in red the links in $\ket{\uparrow}$ 
and not coloring the ones in $\ket {\downarrow}$ that represent the electric states. 
Physical states are superposition of endless red strings going up and right (see panel c and Fig. \ref{fig:duality}).
Such electric configurations correspond to position states and $\hat{\cal S}^\dagger_s +\hat{\cal S}_s$ acts as a kinetic term on them.   
{\bf c)} The square of the plaquette operator $\big(\hat{\cal S}^\dagger_s +\hat{\cal S}_s\big)^2$ is diagonal in the electric basis  and plays the role of a potential term that counts flippable plaquettes. 
The competition of the kinetic and potential terms in the Rokhsar-Kivelson Hamiltonian \eqref{eq:HRK} gives rise to a rich phase diagram, with the resonating valence bond solid (RVBS) phase separating
two crystal phases, the Néel and the columnar phases (see description in the main text). The resonant plaquettes are depicted by diagonal double arrows. The alternated green and pink arrows reflect that
the resonant plaquettes are correlated within the same sublattice and anti-correlated with the ones in the opposite sublattice. 
} 
\label{fig:GM}
\end{center}
\end{figure}

\section{Rydberg gauge theories: Dual formulation in terms of generalized blockades}\label{sec:dual}
In this section we show how to formulate a relevant U(1) gauge theory in terms of interactions 
that are natural for an atomic system. Let us start by introducing the gauge theory in simple terms.
\ac{ We focus on the situation in which the gauge field evolves in a background of static charges 
 (possible extension are discussed in the outlook).}

\subsection{Spin gauge theories and the Rohksar-Kivelson model}\label{sec:GM}
The Hamiltonian of a gauge theory in two (or more) dimensions is constructed in terms of electric and magnetic {interactions} and of their coupling to charges. 
In continuous Abelian U(1) gauge theories relevant in high-energy physics like quantum electrodynamics (and similarly for non-Abelian gauge theories like quantum chromodynamics) 
the former are simply given by the square of the electric and magnetic fields $E_\mu$ and $B$, respectively, 
with $E_\mu=\partial_t A_\mu$ and $B=\partial_x A_y-\partial_y A_x$ defined (in the unitary gauge) through the vector potential $A_\mu$. Here $t,x,y$ are the time and space coordinate in 2D, and $\mu=x,y$.
Gauge invariance, i.e. invariance of the Hamiltonian under local phase (symmetry) transformations of the charges, follows directly from the invariance of $E_\mu$ and $B$ 
under $A_\mu\to A_\mu + \partial_\mu \theta(x,y)$. The electric field is sourced by the charges through the Gauss law, $\partial_\mu E_\mu =4 \pi Q$, where $Q$ is the charge density. 

In gauge theories defined on the lattice \cite{Wilson74}, 
the charges occupy the sites $s =(x_s,y_s)$ of the lattice while the electromagnetic field lives on its links $l$. The links are oriented and can be denoted 
by their starting site and their direction $\mu=x,y$, $l=(s,\mu)$. 
The electric interactions are defined directly in terms of the electric operator $\hat E_{s,\mu}$, a Hermitian operator of discrete spectrum acting on the links. 
For each link one further defines a Wilson operator,  $\hat U_{s',\mu'}$, as the lowering operator for the electric field: $[\hat E_{s,\mu},\hat U_{s',\mu'}] =-\hat U_{s,\mu}\delta_{s,s'}\delta_{\mu,\mu'}$.
The Wilson operator measures the phase acquired by a unit charge moved along the link $(s,\mu)$ of length $a$, i.e. $\hat U_{s,\mu} \sim e^{i a A_\mu(s)}$.
The magnetic interactions are given by (oriented) products of these Wilson operators
on the links around the plaquettes of the lattice.
These operators are used to construct the Kogut-Susskind Hamiltonian \cite{Kogut75}.
In the limit of infinitely massive charges such Hamiltonian contains no dynamics for the charges and reads  
\begin{multline}
H_{KS}= \sum_s\left[ \frac{g^2}2 \left(\hat E_{s,x}^2+\hat E_{s,y}^2\right) \right.\\ 
\left. - \frac 1{2g^2}\left( \hat U_{s,x}^\dagger \hat U_{s+\hat x,y}^\dagger \hat U_{s+\hat y,x} \hat U_{s,y} +H.c.\right)\right], \label{eq:HKS}
\end{multline} 
which reduces to the pure gauge U(1) Hamiltonian in the continuum, $H=\int (E^2+ B^2)$, when the lattice spacing $a$ is send to zero. 
The Hamiltonian \eqref{eq:HKS} is gauge invariant as it commutes with the lattice version of the Gauss law 
\begin{multline}
\left(\hat E_{s,x} + \hat E_{s,y} - \hat E_{s-\hat x,x} - \hat E_{s-\hat y,y}- \hat Q_s\right) \ket{\Phi} =0,\ \ \forall s, \\
\iff \ket{\Phi} \in \text{\{physical states\}}, \label{eq:GaussKS} 
\end{multline}
that determines what states are physical for a given distribution of charges. Here, $\hat Q_s$ is the operator measuring the charge on the site $s$ and 
$\ket{\Phi}$ represents the state of the whole lattice, including both links and sites. 
The electric states, i.e. the eigenstates of the electric operators on the links, form a convenient basis for the link degrees of freedom. 
In particular, the physical states can be easily identified in this basis via \eqref{eq:GaussKS}.   

Since the electric field is unbounded, the number of electric states on each link are in principle infinite.
However, it is possible to truncate it to a maximal value and define consistently U(1) (and SU(N) \cite{Brower99,Tagliacozzo14}) lattice gauge theories
with finite local Hilbert spaces (at the price that the Wilson operator is no longer unitary).
The simplest U(1) gauge theories in 2D, known as gauge magnets, link models, Ising gauge theories \cite{Horn81,Orland90,Chandrasekharan97,Moessner01b},
are obtained by considering just two electric states per link, see Fig. \ref{fig:GM}. 
The electric operator reduces to $\hat E_{s,\mu}\to \hat S^z_{s,\mu}$ and the Wilson operator to $\hat U_{s,\mu}\to \hat S^-_{s,\mu}$, with $\hat S^{\pm} = \hat S^x \pm i \hat S^y$, 
and the physical configurations and their dynamics follow from \eqref{eq:GaussKS} and \eqref{eq:HKS}. 

We can represent the physical configurations in the (electric) $\hat S^z$ basis by coloring in red the links in $\ket{\uparrow}$ and not coloring the ones in $\ket {\downarrow}$.
We use such notation in Fig. \ref{fig:GM}b to illustrate the 6 configurations out of 16 allowed around a site without charges. All the physical configurations of the plane are obtained 
by assembling the local building blocks satisfying \eqref{eq:GaussKS}. 
  
Since $E^2$ is trivial ($(\hat S^z)^2= 1$) \ac{in the retained states  \cite{fnote5}}, the Kogut-Susskind Hamiltonian \eqref{eq:HKS} specialized to the truncated theory contains only magnetic interactions 
$H_{KS}\to - \sum_s \hat {\cal S}_s +\hat {\cal S}_s^\dagger$ with
\begin{equation}
\hat {\cal S}_s +\hat {\cal S}_s^\dagger = \hat S^-_{s,x} \hat S^-_{s+\hat x,y} \hat S^+_{s+\hat y,x} \hat S^+_{s,y} + H.c., \label{eq:plaquette}
\end{equation}
where we label the plaquette by its lower left site. As shown in Fig. \ref{fig:GM}b, 
$\hat {\cal S}_s +\hat {\cal S}_s^\dagger$ interchanges two electric configurations of a plaquette, 
while it annihilates the remaining (fourteen) ones. In the following we refer to these two configurations as \emph{flippable} and to $\hat {\cal S}_s +\hat {\cal S}_s^\dagger$ as plaquette operator . 
Notice that neighbor plaquette operators are not commuting: such property is a consequence of the electric field truncation and has important 
implications on the dynamics. \ac{In fact we can interpret the truncation as the effect of a modified electric term, see App. \ref{sec:truncation}. Thus, the usual dynamical competition 
between electric and magnetic interactions is retained and transformed from a soft into a {\it hard} constraint.}
 
The Hamiltonian considered so far is not the most general one compatible with gauge invariance, i.e. commuting with \eqref{eq:GaussKS}, that one can construct.
If we regard the electric basis as a position basis we can interpret the sum of  plaquette operators as a kinetic term that acts on the electric configurations by interchanging them. 
We can thus add potential terms that are diagonal in the electric basis and weight the different configurations.
The resulting lattice models have a rich phase diagram and are of direct interest in condensed matter even without taking the continuum limit. 
A relevant Hamiltonian in this class is the Rokhsar-Kivelson (RK) model \cite{Rokhsar88}
\begin{equation}
H_{RK} =  -J \sum_s \left[  \left(\hat {\cal S}_s +\hat {\cal S}_s^\dagger\right) - \lambda \left(\hat {\cal S}_s +\hat {\cal S}_s^\dagger\right)^2 \right], \label{eq:HRK}
\end{equation}             
where the potential term is given by the square of the plaquette operator. Here the potential term simply counts the number of flippable plaquettes 
and therefore $\lambda$ plays the role of a chemical potential for the flippable electric configurations. 
The RK Hamiltonian was originally proposed as a simple model of \ac{close-packed quantum dimers that could host short-range resonating valence bond \cite{Pauling49,Anderson73,Fazekas74} insulating states.  
Quantum dimers are an effective description of valence bonds, the singlets of valence electrons in a quantum antiferromagnet. 
Originally, it was thought that in presence of doping (which implies the addition of kinetic terms for dimers in \eqref{eq:HRK}) the resonating valence bond state would become a superconductor due to the condensation of 
the vacancies (holon), such to provide a mechanism for high-T$_c$ superconductivity \cite{Anderson87} in cuprates.
Nowadays, it is still expected that the pseudogap phases at moderate hole doping are resonating valence bond phases although of more complex nature \cite{Chowdhury16},
and the properties of the latter are well understood in terms of emerging gauge theories \cite{Sachdev16}, as first realized in \cite{Baskaran88,Fradkin90}.} 
  Dimer configurations are equivalent to electric configurations in the presence of a staggered distribution of static charges $\pm 1$ 
  and their dynamics is described by \eqref{eq:HRK}, see Appendix \ref{sec:qdimer} for further details.  

\begin{figure*}[htb]
\begin{center}
\resizebox{1.95\columnwidth}{!}{\includegraphics{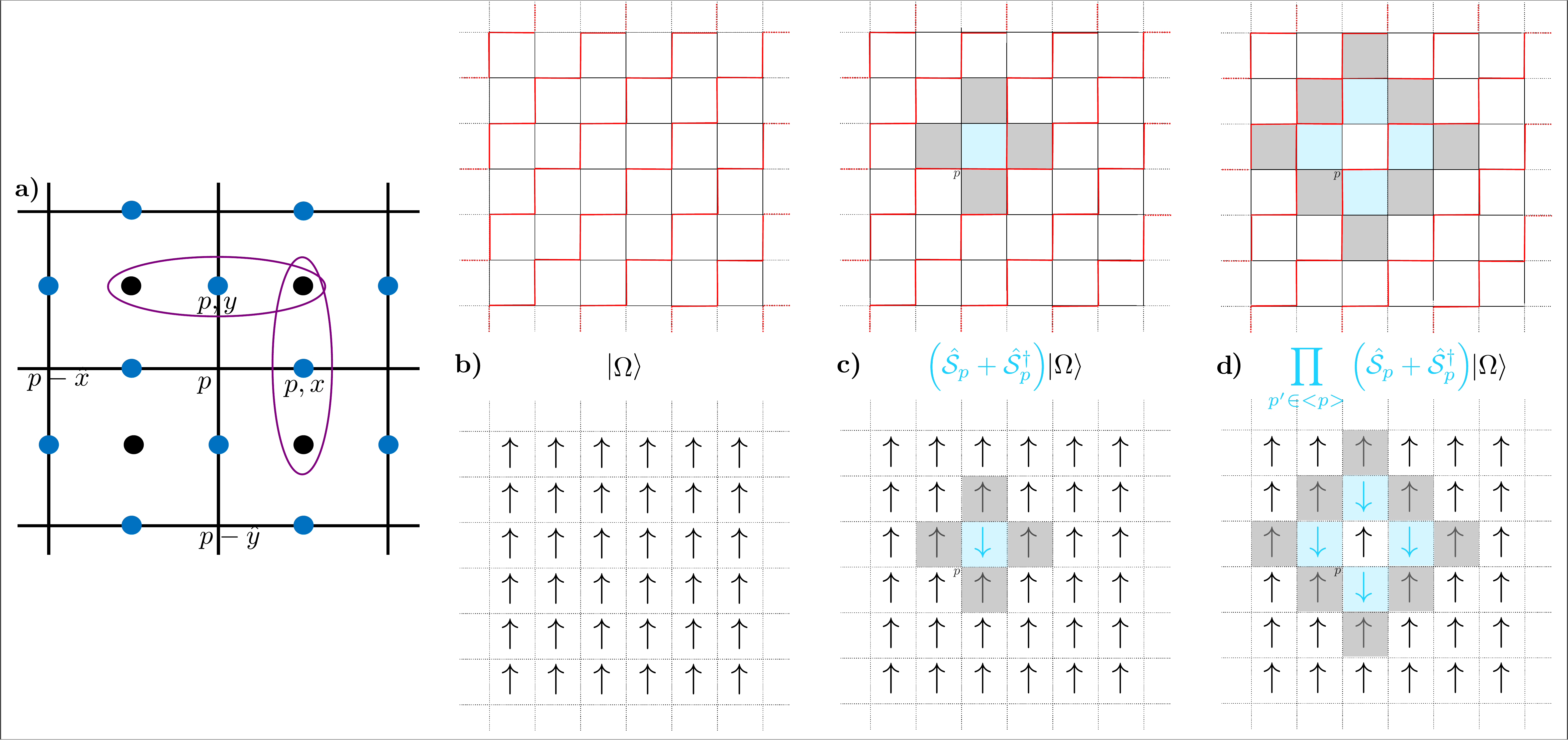}}
\caption{(color online) {\bf Graphical illustration of the duality.} 
{\bf a)} Definition of the link spins (blue dots) in terms of the dual plaquette spins (black dots), \ac{as in} \eqref{eq:formalmap}. The dual spins live in the dual lattice formed by the centers of the plaquettes. 
The value of the link spin is determined by the plaquette spins of the \ac{the two plaquettes sharing the link, as indicated by the purple ellipses for the links $p,x$ and $p,y$.  
The orientation of the plaquette spins are chosen such that the fully flippable state $\ket \Omega$ is mapped in the dual basis into the ferromagnetic state (panel b)).} 
{\bf b), c), d)} We represent \ac{the configurations that are relevant to understand the generalized blockade condition simultaneously in the original (above) and the dual (below) basis.
In the former, we represent the link spins with the color convention of Fig. \ref{fig:GM}. In the latter, we represent the dual plaquette spins
as arrows placed in the center of the corresponding plaquettes.} 
{\bf b)}  The reference state $\ket{\Omega}$: we identify it with the ferromagnetic state with all plaquette spins up. 
  {\bf c)}  After flipping one plaquette of $\ket{\Omega}$ all the neighboring plaquettes are blocked.
{\bf d)} After flipping all \ac{its neighboring plaquettes, the plaquette $p$} is flippable. 
} 
\label{fig:duality}
\end{center}
\end{figure*}

\subsubsection{Phase diagram of the RK model}
The RK Hamiltonian has a rich phase diagram in any lattice geometry and charge distribution. 
At $\lambda=1$, known as the RK point, the Hamiltonian \eqref{eq:HRK} becomes a sum of projectors and is semi-positive definite by construction \cite{Rokhsar88}. 
The equal superposition of all the allowed configurations is the exact, zero energy ground state of $H_{RK}(\lambda=1)$.
This state is a prototype of a quantum spin liquid \cite{Balents10}. 
In the square lattice and with no charges (see Fig. \ref{fig:GM}c), 
the RK point separates a columnar phase for $\lambda>1$, from a resonating valence bond solid (RVBS) phase \cite{Read89,Read90}
with quasi longe-range order that extends 
from $\lambda_c <\lambda<1$.
\ac{For $\lambda=\lambda_c \sim - 0.3$, there is a weakly-first-order phase transition 
\cite{Senthil104,Banerjee13} to a N\'eel phase. 
Both the N\'eel and the RVBS are examples of order by disorder \cite{Villian80}, where quantum fluctuations resolve the classical degeneracy and select a unique ground state. 
The RVBS is also a crystal order phase but richer: similarly to the {\it plaquette phase} \cite{Jalabert91} in the quantum dimer model it preserves the point-symmetry of the lattice
and can be interpreted as the oscillation between the two N\'eel states.
Its correlation pattern (see Fig. \ref{fig:GM}c) spontaneously breaks translational invariance while preserving charge conjugation
\cite{fnote1}.
}

On frustrated lattices like the triangular and kagom\'e ones the corresponding RK Hamiltonians for quantum dimers 
\cite{fnote2}
are expected to display a true spin-liquid phase around the RK point, $\lambda \lesssim 1$,   
 \cite{Moessner01res,Shannon04,Misguich08},
while the overall structure of the phase diagram is lattice dependent, see e.g. \cite[Fig. 17.8]{Moessner11}.

The multi-body interactions in Hamiltonians like the RK model \eqref{eq:HRK} make experimental observations challenging, both in condensed matter systems and in synthetic quantum matter. 
The relevant  Hamiltonian terms can in principle be obtained as low-energy limits of antiferromagnets, for instance in the 2D pyrochlore lattice in the Ising limit \cite{Shannon04}. 
 There, the dominating Ising spin-spin interactions impose the Gauss law on the low-energy manifold and the plaquette interactions emerge 
 in perturbation theory through ring exchange \cite{Buchler05,Glaetzle14}. 
 Alternatively, one can engineer both Gauss law and plaquette interactions by adopting a digital
 approach \cite{Weimer10,Tagliacozzo13}  based on Rydberg gates \cite{Mueller09}.
 In both cases, the suppression of the energy scale/complexity of the digital procedures makes the realization of 2D gauge theories in experiments extremely hard. 

\subsection{Dual formulation of the RK model}\label{sec:duality} 
We show here that it is possible to reformulate spin gauge theories such that the plaquette interactions acquire a simpler form. 
In particular, we find a dual formulation where the multi-body interactions have a natural realization in atomic arrays with Rydberg interactions. 
For the sake of concreteness, let us focus on the RK model on the square lattice without charges 
(and fix the boundary conditions compatible with the N\'eel state), see Figs. \ref{fig:GM}c and \ref{fig:duality}. 
For the dual formulation, we consider a spin-1/2 system associated with each plaquette of the square lattice. 
The physical states of the original gauge theory are related to states of the plaquette spins via the the operator identification
\begin{align}
\hat S^z_{p,x}&\to - 2(-1)^p \hat S^z_{p} \hat S^z_{p-\hat y}\cr
\hat S^z_{p,y}&\to   2(-1)^p \hat S^z_{p} \hat S^z_{p-\hat x}. \label{eq:formalmap}
\end{align}
where the $\hat S^z_p$ acts on the spin 1/2 associated to the plaquette $p=(x_p,y_p)$, and $(-1)^p=(-1)^{x_p+y_p}$ distinguishes even (+) and odd (-) plaquettes.   
Equation \eqref{eq:formalmap} defines a one-to-one mapping between Hilbert spaces that is well defined up to an overall $Z_2$ identification of the plaquette spins for periodic boundary conditions.
For open lattices, the $Z_2$ degeneracy can be removed for instance by choosing all the dual spins up on the boundary. Note that the transformation \eqref{eq:formalmap}
maps the N\'eel state into the ferromagnetic states, 
$\ket{\Omega} = \prod_{p\in \text{ even}} \ket{\color{red}\text{$\halfboxup{\phantom{1}}$}}_p  \to \prod_p \ket{ \uparrow}_p,$
see Fig. \ref{fig:duality}a.\\

This mapping is appealing, because the Hamiltonian \eqref{eq:HRK} is particularly simple in this dual formulation. Since the map \eqref{eq:formalmap} is quasi-local, it maps local operators into local operators. 
In particular, as detailed in appendix \ref{sec:formal}), the plaquette operator can be written as
\begin{equation}
\hat {\cal S}_p +\hat {\cal S}_p^\dagger\to \Big(P^{\uparrow\uparrow\uparrow\uparrow}_p\, +\,P^{\downarrow\downarrow\downarrow\downarrow}_p  \Big)
2 \hat S_p^x,\label{eq:map2D}
\end{equation}
Here, $P^{\uparrow\uparrow\uparrow\uparrow}_p \equiv \prod_{p'\in\langle p \rangle}(\frac 12 + \hat S^z_{p'})$ denotes the projector onto states where all spins neighboring to $p$ are in the state $\ket{\uparrow}$ 
(and analogously $P^{\downarrow\downarrow\downarrow\downarrow}_p$).  
Even though the plaquette operator is a multi-spin operator also in this dual representation, it has a simple form: it  flips the spin of the associated plaquette, conditional on the state of its nearest neighbors. 
In particular, a single plaquette spin is flipped by the plaquette operator only if all four spins on the neighboring plaquettes point in the $\uparrow$-direction or if all point in the $\downarrow$-direction. 
The (dual) RK Hamiltonian is simply given by a sum of such terms
\begin{equation}
H^*_{RK} = - J \sum_p \Big(P^{\uparrow\uparrow\uparrow\uparrow}_p\, +\,P^{\downarrow\downarrow\downarrow\downarrow}_p  \Big)\left(2 \hat S^x_p -\lambda\right). \label{eq:dualHRK2D}
\end{equation}
Note that the map \eqref{eq:formalmap} and the Hamiltonian \eqref{eq:dualHRK2D} are the quantum version of the height formalism \cite{Sachdev89,Levitov90,Henley97} applied to RK model without charges, see appendix \ref{sec:formal}. 
The same map can be derived also for other charge sectors and geometries \cite{Celi19}, including frustrated lattices 
\cite{fnote3}
(see \ref{sec:outlook} for further discussion).

\begin{figure*}[htb]
\begin{center}
\resizebox{1.9\columnwidth}{!}{\includegraphics{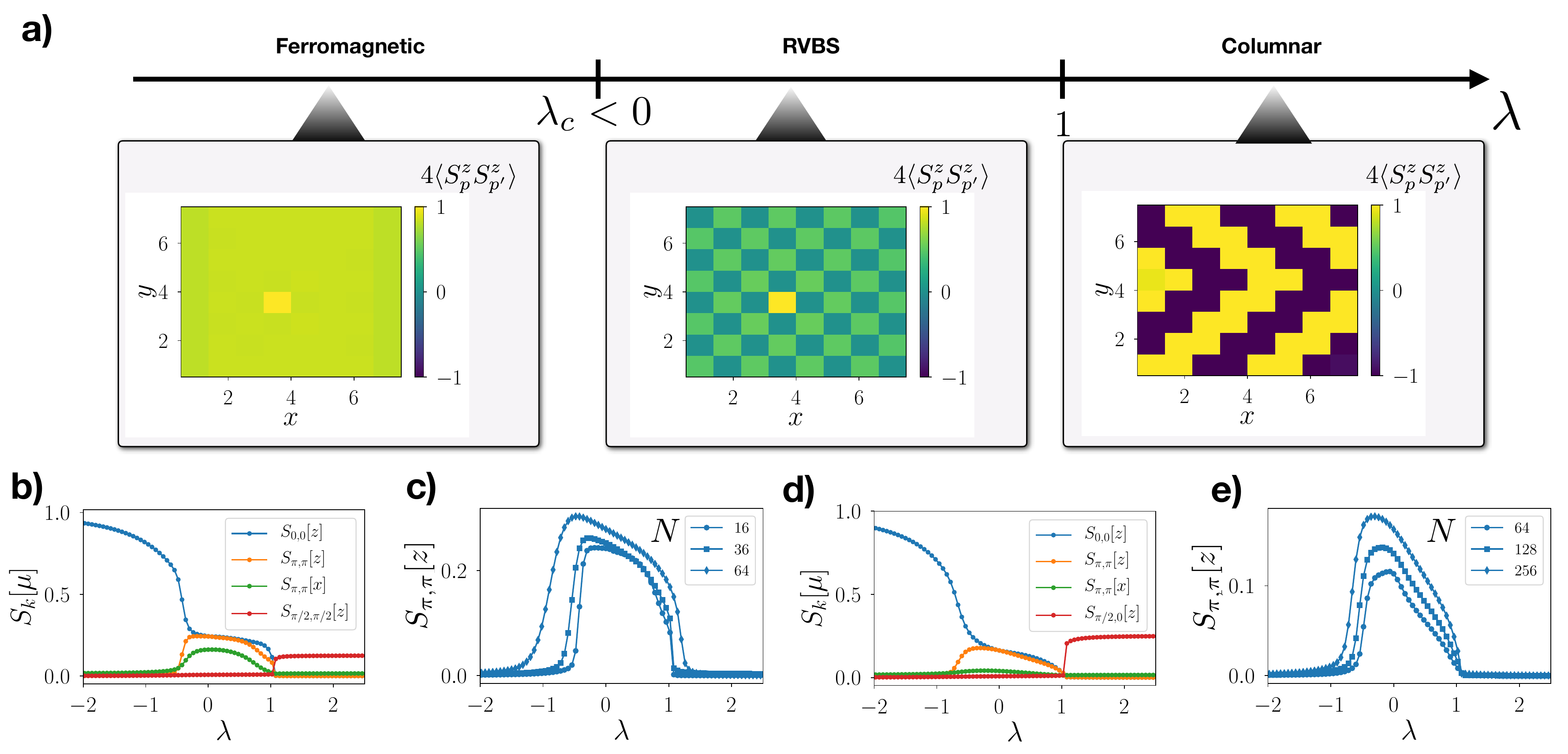}}
\caption{(color online) {\bf Dual RK model and its phases.}
{\bf a)} Phase diagram of the 2D RK model on the square lattice without background charges. In the boxes we plot the spin-spin correlations $\langle S_p^\mu S_{p'}^\mu \rangle$ 
obtained via DMRG \cite{ITensors} on square lattices $N_x\times N_y=8\times 8$, and for three values of $\lambda=-1,0,2$ that belong to the ferromagnetic (Néel), RVBS, and columnar phases, respectively. 
The correlations are represented with respect to the center, $p=(4,4)$, for varying $p'$.
Note that in the $\lambda \to \infty$ limit there are multiple degenerate unflippable states. The particular configuration shown here in the right panel corresponds to one of them. 
{\bf b-c)} Structure factors $S_k[\mu]= 4/(N_xN_y)^2\sum_{p,p'}\exp[i (p-p')k]\langle \hat S^\mu_p \hat S^\mu_{p'}\rangle$, $\mu=x,z$ for $N_x\times N_y=8\times 8$ [panel (b)], 
and for different system sizes $N=N_x N_y=N_x^2$ [panel (c)]. The raising of the structure factors  $S_{(\pi,\pi)}[x,z]$ identifies the RVBS phase. In particular, $S_{(\pi,\pi)}[z]$ equals $S_{(0,0)}[z]$ in the RVBS region.
{\bf d-e)} Structure factors on the periodic ladder of $N_x\times 2$ sizes, without background charges, with $N_x=32$ [panel (d)], and  $N_x=32,64,128$ [panel (e)].} 
\label{fig:dRK}
\end{center}
\end{figure*}

To illustrate this duality further, let us consider the repeated action of the plaquette operator on the fully flippable state $\ket{\Omega}$. 
Once the first plaquette operator $(\hat {\cal S}_p +\hat {\cal S}_p^\dagger)$ is applied, all spin on plaquettes $p'$ neigbouring  $p$ can no longer be flipped. 
This reflects the gauge constraint (and the electric-field truncation) and manifests in the annihilation of the state by the corresponding plaquette operator, 
$(\hat {\cal S}_{p'} +\hat {\cal S}_{p'}^\dagger)(\hat {\cal S}_p +\hat {\cal S}_p^\dagger)\ket{\Omega}=0$ (see Fig. \ref{fig:duality}c). 
In the context of Rydberg physics, such conditional or constraint dynamics is well known as blockade effect.  
However, in contrast to the standard blockade mechanism, where a spin can be flipped only if all its neighbors are in a single, specific configuration, 
here two distinct configurations of its neighbors allow a spin to flip. 
We thus refer to the present situation as generalized blockade condition. 
Specifically, if all of the remaining three spins on plaquettes neighboring to $p'$ are flipped, 
then the spin $p'$ becomes flippable again, see Fig. \ref{fig:duality}d. 
It is easy to see that one can in fact span all physical states --the states that satisfy the gauge constraint-- (and no other states) by repeated application of plaquette operators.

\subsubsection{Phase diagram of the RK model in the dual formulation}

We conclude by discussing the phase diagram and in particular the RVBS phase in the dual picture 
\eqref{eq:dualHRK2D}, see Fig. \ref{fig:dRK}.
There, we show also the corresponding structure factors $S_k[\mu]=\sum_{p,p'} 4/(N_xN_y)^2\exp[i (p-p')k]\langle \hat S^\mu_p \hat S^\mu_{p'}\rangle$ 
calculated by DMRG using ITensor library \cite{ITensors} on square lattices $N_x\times N_y$ up to $8\times 8$.
We ensure that the energy is minimized only on accessible states from $\ket{\Omega}$, for which the map is defined (see Appendix \ref{sec:DMRG}).   
The phase for $\lambda<\lambda_c$ is ferromagnetically ordered in each sublattice as expected for the fully flippable phase. 
The boundary conditions compatible with $\ket{\Omega}$ select the ferromagnetic order in both sublattices as evidenced by a large value of $S_k[z]$ for $k=(0,0)$, the blue curve in Fig.  \ref{fig:dRK}b, 
which is the dominant contribution to spin-spin correlations.
The RVBS for $\lambda_c<\lambda<1$ is correlated both along $z$ and $x$ within the two sublattices only. Such structure is in agreement with RVBS being even under charge conjugation and
it is signaled both by the equality of the $S_k[z]$ for $k=(0,0)$ and $k=(\pi,\pi)$ (in orange) and the raising of the $S_k[x]$ (in green), see Fig. \ref{fig:dRK}b-c.
The ``unflippable'' phase (it is reachable from $\ket \Omega$ only through the boundary where the residual flipplable plaquettes appear as defects) for $\lambda>1$ displays a characteristic strip order with the plaquette spins aligning in the $z$ direction along the diagonals of the lattice, with an associated periodicity four captured by $S_k[z]$ for $k=(\pi/2,\pi/2)$. In Fig. \ref{fig:dRK}a we show one of the possible configurations with two flippable plaquettes (defects) on the boundary. For $\lambda\to \infty$ we have as many exactly degenerate configurations as allowed positions for the defects. The degeneracy is only approximated at finite $\lambda>1$ (the degeneracy is resolved): in 
Fig. \ref{fig:dRK}a we show the resulting configuration.

\subsubsection{Special cases: ladder geometries}
The dual RK Hamiltonian \eqref{eq:dualHRK2D} is reminiscent of Rydberg physics and of the $PXP$ model \cite{Fendley04,Bernier17,Samajdar18}, 
which can be obtained from \eqref{eq:dualHRK2D} by replacing the projector $P^{\uparrow\dots\uparrow}_p\,+\, P^{\downarrow\dots\downarrow}_p$ with the simpler blockade condition
 $P^{\uparrow\dots\uparrow}_p$ (or $P^{\downarrow\dots\downarrow}_p$).  
It was noticed in \cite{Moessner01} that the quantum dimer model on a ladder can be mapped into a $PXP$-like model on a chain.
Indeed, if we restrict the Hamiltonian \eqref{eq:dualHRK2D} to a chain immersed in the background $\ket{\Omega}$,  
only the blockade term $P^{\uparrow\uparrow}_p=(\frac 12 + \hat S_{p-\hat x}) (\frac 12 + \hat S_{p+\hat x})$ survives. 
Precisely, the resulting Hamiltonian coincides with \cite[Eq.3]{Fendley04} for $\omega=J$ and $U=-2V=-2\lambda J$.
It is not charge conjugation invariant and the corresponding phase diagram for varying $\lambda$ 
does not show any analogue of the RVBS phase, which is even under charge conjugation,  
see Fig. \ref{fig:1D} in the Appendix \ref{sec:PXP}.  

Interestingly, the minimal geometry that preserves such symmetry is a periodic ladder, i.e. a cylinder with a circumference of two lattice sites. 
Since the dual lattice of a periodic ladder is a periodic ladder, 
$p+\hat y=p-\hat y$, the dual RK Hamiltonian takes the simple form
\begin{equation}
H^{*pl}_{RK} = - J \sum_p \Big(P^{\uparrow\uparrow\uparrow}_p + P^{\downarrow\downarrow\downarrow}_p \Big)\left(2 \hat S^x_p -\lambda\right), \label{eq:dualHRKpl}
\end{equation}
where we restrict $p$ to run over the sites of a square ladder, $p=(x_p,y_p)$ with $y_p=[0,1]$, such that 
$P^{\uparrow\uparrow\uparrow(\downarrow\downarrow\downarrow)}_{x,y} = (\frac 12 +(-) \hat S^z_{x-1,y})(\frac 12 +(-) \hat S^z_{x,y+1})(\frac 12 +(-) \hat S^z_{x+1,y})$. 
As shown in Fig. \ref{fig:dRK}d-e, the phase diagram of \eqref{eq:dualHRKpl} has the same structure as the one of the 2D RK model,  Fig. \ref{fig:dRK}b-c. 
In particular, the RVBS phase is signalled by the equality of the structure factors $S_k[z]$ for $k=(0,0)$ and $k=(\pi,\pi)$. 

\begin{figure*}[htb]
\begin{center}
\resizebox{1.95\columnwidth}{!}{\includegraphics{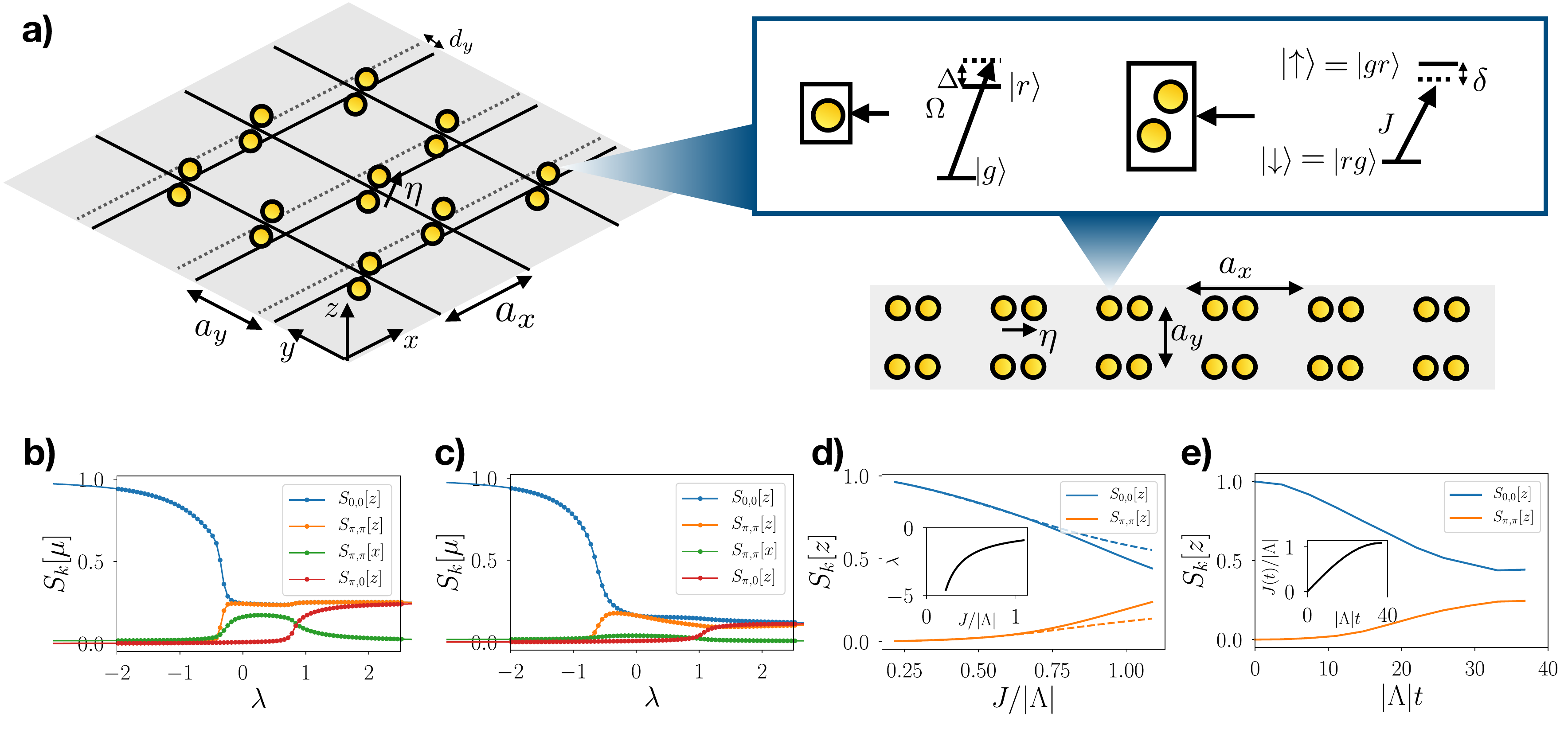}}
\caption{(color online) {\bf RVBS in decorated Rydberg arrays.} 
{\bf a)} Implementation of the RK Hamiltonian in decorated Rydberg arrays: oriented pairs of closeby atoms with blue detuned Rabi oscillations between the ground $\ket g$ and Rydberg $\ket r$ states
are equivalent to oscillating spins 1/2, see inset, with Ising interactions between them. By controlling the pair orientation relative to the lattice and the lattice geometry we can engineer the spin-spin interactions
such to achieve the generalized blockade conditions and realize the RK Hamiltonian in the dual formulation with a modified RK potential. We name such Hamiltonian as Rydberg-RK Hamiltonian. 
On the left we show the square lattice construction and on the bottom right the implementation for a ladder. 
The same principle applies to a generic 2D lattice.  
{\bf b-c)} The phase diagrams of the Rydberg RK Hamiltonian on the square lattice \eqref{eq:Heffspinsquare} [$N=8\times8$, panel (b)] and  on the (periodic) ladder \eqref{eq:Heffspinpl} [$N=32\times2$, panel (c)] 
as a  a function of $\lambda$, calculated via DMRG.
 The two phase diagrams differ qualitatively from the ones of the original RK Hamiltonians shown in Fig. \ref{fig:dRK}b and c, respectively, 
for $\lambda \gtrsim 1$
In particular, the ferromagnetic and the RVBS phases remain while the columnar phase is substituted by a ``glassy'' phase.  
{\bf d)} In order to compare the phase diagrams of the effective RK Hamiltonian with the full spin description of the decorated Rydberg array, 
we compare the spin-spin correlations on a $6\times 2$ decorated ladder, for $|\eta|=0.38a_x$ and periodic boundary conditions, via exact diagonalization. 
The structure factors for the ground state of the effective RK Hamiltonian, and the corresponding eigenstates of the Rydberg array Hamiltonian (up to next-nearest-neighbor interactions) 
are shown respectively as dashed and solid lines. We included a small detuning $\delta=0.1C_6/a^6_x\approx 0.11|\Lambda|$ within the atoms of each pair to select a specific ferromagnetic state as reference.
{\bf e)} Adiabatic preparation of the RVBS phase for the same parameters as d).  
By smoothly ramping up the Rabi frequency $J$ (inset) we can enter the RVBS phase  within the validity of generalized blockade condition, $J\ll G$.  
} 
\label{fig:imple}
\end{center}
\end{figure*}

\section{Rydberg gauge theories: RK Hamiltonian from decorated Rydberg arrays}\label{sec:imple}
In this section we discuss the implementation of the dual RK Hamiltonian \eqref{eq:dualHRK2D} (and of analogous expression for other lattices) with Rydberg atoms.
Our starting point is a 2D array of atoms with tunable geometry, driven between their ground $\ket g$ and Rydberg $\ket r$ states, respectively, with Rabi frequency $\Omega$ and  
detuning $\Delta$ ($\hbar=1$) 
\begin{multline}
H_{Ryd} =\sum_I\left[ -\Omega \big(\ket{r}_{I}\bra{g} +\ket{g}_{I}\bra{r}\big) + \Delta\, n_I \right.  \\
               \left.+ \frac 12 \sum_{I'\neq I} \frac {C_6}{|I-I'|^6} n_I n_{I'}\right],\label{eq:Hryd}
\end{multline} 
where $n_I=\ket{r}_{I}\bra{r}$ and $C_6$ is the van der Waals potential energy at unit distance.
We arrange the atoms by pairs such to realize composite spins 1/2.

We first show in Sec. \ref{sec:tunable_Ising} that under suitable conditions $H_{Ryd}$ becomes equivalent 
to an anisotropic Ising model for the atom pairs with interactions and detuning determined by the pair arrangement.
Then, in Sec. \ref{sec:periodic_ladder} and \ref{sec:2Dgenblock} we achieve the generalized blockade from such spin model 
by properly choosing the pair arrangement in a similar fashion as for ordinary blockade in Rydberg chains \cite{Bernier17}.

\subsection{Tunable Ising models from decorated Rydberg arrays} \label{sec:tunable_Ising}
We consider an array of atom pairs (see Fig. \ref{fig:imple}) such that the position of each atom $I$ is given in terms of the pair position $p$ and atom displacement in the pair $\eta$,  
$I=p+\sigma \eta$, $\sigma=\pm\frac 12$. We take $|\eta|$ sufficiently small (i.e. the pairs sufficiently far apart)  
such that Rydberg interactions between the atoms 
in the pairs is the dominating interaction, $\frac {C_6}{|\eta|^6}\gg \Omega,\frac {C_6}{|I(p)-I(p')|^6}$, $p'\neq p$. 
For a sufficiently blue detuned driving field ($\Delta<0$ such that $\Omega\ll -\Delta$),
the low energy sector consists of configurations where exactly on atom per pair is in the Rydberg state and the other atom is in the internal ground state. 
Thus, each pair forms an effective pseudo spin-1/2 system, with the identification (see the inset of Fig. \ref{fig:imple}a)  
\begin{equation}
  \hat S_p^z \ket{g}_{p-\sigma\eta}\ket{r}_{p+\sigma\eta} =\sigma \ket{g}_{p-\sigma\eta}\ket{r}_{p+\sigma\eta}.\label{eq:effspin}
  \end{equation} 
The effective Hamiltonian \eqref{eq:Hryd} in this low-energy limit becomes
\begin{equation}
 H_{Ryd} \approx -2 J \sum_p \hat S_p^x + \sum_k V^{(k)}, 
 \label{eq:Heffspin}
 \end{equation}
where $J$ is the effective Rabi frequency, $J =\Omega^2 (\frac 1{\Delta} + \frac{|\eta|^6}{C_6-\Delta |\eta|^6})$, and  
$V^{(k)}$ is the van der Waals interactions between Rydberg atoms in $k-$neighbor pairs. 
Disregarding boundary and constant terms, $V^{(k)}$  can be written as Ising interactions between the $k$-spins plus a local term (such term appears because the distance between the atoms in the Rydberg state 
in the pair-pair configurations $\uparrow_p\downarrow_{p'}$ and $\downarrow_p\uparrow_{p'}$ can differ)     
\begin{equation}
V^{(k)}=\sum_p \hat S^z_p \sum_{p'=\langle p \rangle_k} \bigg[A(p-p',\eta)\hat  S^z_{p'} +  B(p-p',\eta)   \bigg],\label{eq:Vk}
\end{equation}  
where $\langle p \rangle_k$ denotes the $k$-neighbors of $p$. 
The coefficients $A$ and $B$ are determined by the geometry of the array: they are symmetric and antisymmetric functions of the pair separation and of the atom displacement in the pairs $\eta$
\begin{align}
A(p,\eta)&= C_6 \left(\frac 2{|p |^6} - \frac 1{|p + \eta|^6} - \frac 1{|p -\eta|^6}\right), \label{eq:Apeta}\\
B(p,\eta)&= C_6 \left(\frac 1{|p + \eta|^6} - \frac 1{|p -\eta|^6}\right). \label{eq:Bpeta}
\end{align}  
Below we show how to exploit their tunability for achieving the dual RK Hamiltonian for various lattice geometries.
In such construction the pair-pair interactions $V^{(k)}$ for $k>2$ are negligible and we consider them zero.

\subsection{RVBS phase on a decorated Rydberg ladder} \label{sec:periodic_ladder}

As a preliminary exercise we consider the implementation of the RK Hamiltonian on a periodic ladder \eqref{eq:dualHRKpl}. 
Here we parametrize the position of the pairs on the ladder as $p=(a_x x_p, a_y y_p)$, where $a_x$ and $a_y$ are the lattice spacing along the legs and the length of the rungs, respectively, see Fig. \ref{fig:imple}a.
For simplicity, we can express all lengths in unit of $a_x$, $a_x=1$.

Let us focus on the projector appearing in \eqref{eq:dualHRKpl}.
We can obtain it dynamically from the nearest-neighbor (NN) pair interactions $V^{(k=1)}$ in \eqref{eq:Vk} if we require that 
$\sum_{p'=\langle p \rangle_1} A(p-p',\eta)=0=\sum_{p'=\langle p \rangle_1} B(p-p',\eta)$.
By orienting the pairs along $x$, $\eta= |\eta| \hat x$, the constant terms $B$'s cancels out. 
The requirement on the NN Ising couplings reduces to $2 A(\hat x,\eta)=- A(a_y \hat y,\eta)\equiv-2G(|\eta|)$, 
which fixes the lattice spacing $a_y$ (more precisely, the ratio $a_y/a_x$) in terms of $|\eta|$. 
With this choice the Hamiltonian \eqref{eq:Heffspin} of a decorated Rydberg ladder becomes 
\begin{align}\label{eq:Heffspinpl}
H_{Ryd} &= V^{(2)}
-\sum_p\left[ 2J \hat S^x_p +G \hat S^z_p (\hat S^z_{p+\hat x} +\hat S^z_{p-\hat x} - 2 \hat S^z_{p+\hat y})\right]
\nonumber\\
       &\hspace{-2mm}\stackrel{
       J\ll G
}{\approx} V^{(2)}-2J \sum_p \Big(P^{\uparrow\uparrow\uparrow}_p + P^{\downarrow\downarrow\downarrow}_p \Big) \hat S^x_p, \hspace{-6mm}
\end{align}
as the non-flippable configurations are separated from the flippable ones by a non zero gap that is precisely equal to $G$. 
Notice that the next-to-nearest neighbor (NNN) pair interactions $V^{(2)}$ can be also written in terms of projectors over NN spins.
For  $\eta= |\eta| \hat x$, $V^{(2)}$ reduces to 
\begin{equation}
V^{(2)} = \sum_p \Lambda(|\eta|) \big( P_p^{\uparrow\uparrow\uparrow} + P_p^{\downarrow\downarrow\downarrow} - P_p^{\uparrow\downarrow\uparrow} - P_p^{\downarrow\uparrow\downarrow} \big),\label{eq:V2ladder}
\end{equation}
with $\Lambda(|\eta|)= A(\hat x + a_y(|\eta|), \eta)/2<0$. Thus, for $J\ll G$, the effective spin Hamiltonian \eqref{eq:Heffspinpl} is equivalent to the RK Hamiltonian with a generalized RK potential.
We name it as the ``Rydberg RK'' Hamiltonian on a periodic ladder. 

As the original RK Hamiltonian \eqref{eq:dualHRKpl}, the Rydberg RK Hamiltonian \eqref{eq:Heffspinpl} displays a RVBS phase that is experimentally accessible.
We plot the phase diagram of \eqref{eq:Heffspinpl} for varying $\lambda= \Lambda/J$,  in Fig. \ref{fig:imple}b for the square lattice, and in Fig.~\ref{fig:imple}c for the periodic ladder, 
and we compare it with the one of \eqref{eq:dualHRKpl}.
The extra potential term in \eqref{eq:Heffspinpl} modifies appreciably the phase diagram only for $\lambda \gtrsim 1$, {\it cf.} with  Fig. \ref{fig:dRK}c. 
In particular, the ferromagnetic and the RVBS phases remain unchanged, as clearly evidenced by the behavior of the structure factor. The $S_{(0,0)}[z]$ dominates in the ferromagnetic phase and coincides with
$S_{(\pi,\pi)}[z]$ in the RVBS phase. For $\lambda>1$ the columnar phase is substituted by a ``glassy'' phase, with a huge classical degeneracy imposed by $V^{(2)}$. 
In the 2D case, the phase transition from the ferromagnetic phase to the RVBS phase moves slightly to the right, $\lambda_c$ from $\sim -.6$ to $\sim -.5$, 
as the additional potential term stabilizes the ferromagnetic phase for $\lambda<0$. The phase transition from the RVBS phase to the glassy phase still occurs around $\lambda= 1$, 
but the model at $\lambda=1$ is no longer integrable.
In Fig.~\ref{fig:imple}c, we test numerically the validity of the mapping from the Rydberg Hamiltonian Eq.~\eqref{eq:Heffspin} to the effective Rydberg RK Hamiltonian \eqref{eq:Heffspinpl} for increasing values of $J$. 
For this, we consider a periodic ladder of $N_x=6 \times 2$ atom pairs, with periodic boundary conditions along the $x$-axis, 
and compare via exact diagonalization the ground state of Eq.~\eqref{eq:Heffspinpl} with the corresponding eigenstate of  Eq.~\eqref{eq:Heffspin}. 
We fix the displacement of the atoms in the pairs to be $\eta=0.38 a_x \hat x$ and the distance of the pairs along $y$ is $a_y\approx 0.59 a_x$.
With such values, we obtain $\Lambda=-0.91 C_6/a^6_x$ and $G=15.71 C_6/a^6_x$.  
 We also consider a pinning Hamiltonian $\delta \sum_p S_p^z$, with $\delta=0.1C_6/a^6_x$ to compare eigenstates of the two Hamiltonians consistently in the presence of degeneracies 
 (in particular for $J\to0$, i.e., $\lambda\to -\infty$). 
 The structure factors $S_k[z]$ shown in Fig.~\ref{fig:imple}c show that the RVBS can be accessed within the regime of validity $J\ll G$  of Eq.~\eqref{eq:Heffspinpl}. 
 In Sec. \ref{sec:prepa}, we show how to form such state via adiabatic state preparation in an experiment.

\subsection{Rydberg gauge theory on the square and other 2D lattices} \label{sec:2Dgenblock}

We now apply the same strategy explained above to engineer the 2D RK Hamiltonians defined on generic lattice geometries, e.g. on triangular, square, and hexagonal lattices.
Their dual formulation respectively on the dual hexagonal, square, and triangular lattices can be written in a unified fashion \cite{Celi19} as
\begin{equation}
H^*_{RK} = - J \sum_p \Big(P^{\uparrow\dots\uparrow}_p\, +\,P^{\downarrow\dots\downarrow}_p  \Big)\left(2 \hat S^x_p -\lambda\right), \label{eq:unifieddualHRK2D}
\end{equation} 
where the projectors $P^{\uparrow\dots\uparrow(\downarrow\dots\downarrow)}=\prod_{p'=\langle p \rangle_1}(\frac 12 +(-) \hat S^z_{p'}) $ involve the three, four, and six NN spins to the spin at the site $p$, respectively.
In order to engineer the generalized blockade conditions associated with such projectors, we exploit  the NN Ising interactions arising from $V^{(1)}$, as in the example of the periodic ladder.
While the dynamical implementation of the projectors for dual hexagonal lattice follows precisely the same path as for the periodic ladder
, the higher lattice coordination of square and triangular lattices introduces new conditions. 
In the latter case, it is not sufficient to ask that $\sum_{p'=\langle p\rangle_1} A(p-p',\eta)=0=\sum_{p'=\langle p \rangle_1} B(p-p',\eta)$ as additional degeneracy can arise. 
For instance, if the Ising couplings to two neighbor spins at $p_1$ and $p_2$ are opposite,  
$A_{p-p_1}=-A_{p-p_2}$, we are free to flip these two spins without paying any interaction energy. Thus, the Ising interactions project at low energies also on additional unwanted configurations.
In practice, in lattices with more than three neighbors we have to design the Ising couplings $A(p-p')$, $p'=\langle p \rangle_1$,  such that their sum for any subset $\{\langle p \rangle_1\}$ of NN spins is non zero. 
The gap $G$ is the modulus of the smallest sum, $G=\text{Min}[|\sum_{p'\in\{\langle p \rangle_1\}}A(p-p',\eta)|]$.

For concreteness, we sketch here how to find the proper array of atoms that
avoids the unwanted degeneracies and realize the Rydberg RK Hamiltonian on the square lattice.
The details of the calculation are presented in the Appendix \ref{sec:impledet}. For the construction on other lattices we refer the reader to \cite{Celi19}.   
In what follows, we consider a rectangular lattice of pairs deformed by the relative displacement of its even and odd sublattices, 
indicated in blue and red in the right panel of Fig. \ref{fig:imple}a. As shown in this figure, we displace the sublattices along $y$ by $d_y \,\hat y$ while the pairs lie in the $xz$-plane, 
with the relative displacement of the atoms parametrized as $\eta=|\eta| (\cos \theta \hat x + \sin \theta \hat z)$.
We fix the lattice spacing along $x$ to $a_x=1$ while the one along $y$ is $a_y$.
For any $\theta$, $d_y$, and $a_y$, the constant detunings in $V^{(1)}$ add up zero. 
The Ising couplings to the left and right NN spins are the same and positive
while for $d_y\neq 0$ the Ising couplings to the top and to the bottom NN spins are different and negative.
By adjusting $a_y$ we can make the sum of the four couplings zero while any sum of two or three of them is not,
and thus achieve the desired blockade condition with a finite gap $G$ to the unflippable configurations.

The effective Hamiltonian takes the RK form 
\begin{equation}
H_{Ryd} \stackrel{J\ll G}{\approx} -2J \sum_p \Big(P^{\uparrow\uparrow\uparrow\uparrow}_p + P^{\downarrow\downarrow\downarrow\downarrow}_p \Big) \hat S^x_p + V^{(2)},
\label{eq:Heffspinsquare}
\end{equation}
with the generalized RK potential determined by NNN interactions 
\begin{equation}
V^{(2)} = \sum_p \Lambda \big( P_p^{\uparrow\uparrow\uparrow\uparrow} + P_p^{\downarrow\downarrow\downarrow\downarrow} 
- P_p^{\uparrow\downarrow\uparrow\downarrow} - P_p^{\downarrow\uparrow\downarrow\uparrow} \big),\label{eq:V2square}
\end{equation}
where we order the NN spins anticlockwise and $\Lambda=\Lambda(\eta,d_y)$.

We can use $d_y$ and $\theta$ to both maximize the gap $G$ and minimize the ratio $|\Lambda/G|$. 
In particular, we can reduce the latter arbitrarily such that the RK Hamiltonian \eqref{eq:Heffspinsquare} is valid for $J\gtrsim \Lambda/\lambda_c$,
and the driven Rydberg array supports a RVBS phase.

\subsection{Adiabatic preparation of the RVBS phase in decorated Rydberg arrays}\label{sec:prepa}

Above we have shown how to engineer Rydberg arrays that naturally realize dual RK Hamiltonians on different lattice geometries. 
In particular, we have shown that the dual RK on a square lattice displays a RVBS phase.
Here we discuss how to prepare this phase and how to access the full phase diagram of such Hamiltonians adiabatically in current Rydberg experiments. 
The idea is to exploit the detuning of the effective Rabi coupling for the 
spins as done in several experiments \cite{Schauss15,Labuhn16,Bernier17,Guardado18,Lienhard18}
. We consider a driven array of atom pairs as in \eqref{eq:Hryd} where the ground states of the two atoms in the pair have an energy 
off-set $\delta$ much smaller than both the detuning and the van der Waals interaction in the pair, $\delta\ll-\Delta,C_6/|\eta|^6$, as produced for instance by an optical lattice.
In the spin language the net effect of such term is to induce a detuning for the spins such that their effective Hamiltonian in generic geometry reads
\begin{align}\label{eq:Heffspindelta}
 H_{Ryd} &\approx  \sum_p \left(-2 J\hat S_p^x + \delta \hat S_p^z\right)  + V^{(1)} + V^{(2)}\\
         &\approx \sum_p \left(-2J \Big(P^{\uparrow\dots\uparrow}_p + P^{\downarrow\dots\downarrow}_p \Big) \hat S^x_p + \delta \hat S_p^z \right)+ V^{(2)}, \nonumber
  \end{align}        
where in the second line we specialize to arrays that satisfy the geometric requirements discussed in Sec. \ref{sec:periodic_ladder} 
and \ref{sec:2Dgenblock} and take the effective Rabi coupling sufficiently smaller than the gap $G$, $J\ll G$.
The presence of the detuning does not change such geometric requirements as it is a diagonal term that commutes with $V^{(1)}$. 
This also implies that we can prepare adiabatically the ground state (or the maximally excited state) of the emergent dual RK Hamiltonian \eqref{eq:Heffspindelta} for $\delta=0$ in any phase accessible within its regime 
of validity.

For concreteness, let us focus on the preparation of the RVBS phase on the square lattice. 
We consider the possibility to vary $J(t)$, for instance via the Rabi frequency $\Omega(t)$, as a function of time.
The idea is then to access the phase from the ferromagnetic phase that is connected to the product state 
with all the spin ups, with $J(t=0)\ll |\Lambda|$. The value of $J(t)$ is then slowly increased to reach the final state of the adiabatic state preparation. 
The procedure is illustrated in the case of the periodic ladder in Fig.~\ref{fig:imple}e, with a pulse $J(t)=J\sin[\pi t/(2t_f)]$, $t_f=40/J$,  shown in inset. 
We simulate the dynamics within Eq.~\eqref{eq:Heffspin}, and for the same parameters as Fig.~\ref{fig:imple}d. For the chosen value of $J=C_6/a^6_x$, the system acquires a RVBS pattern at the final time $t_f$. 

 We expect the protocol to be able to form RVBS states also for larger ladders that are beyond the capability
  of classical computation. In realistic experimental setups, the adiabatic preparation is limited by the finite coherence time of the array due to spontaneous decay of the Rydberg states and to their motion. 
  The inverse of the coherence time sets a minimal speed at which the parameters can change. Such minimal speed sets the minimal energy gap (at the transition between the ferromagnetic and the RVBS phases) compatible with adiabaticity. 
  Thus, the coherence time determines the maximal size of the ladder whose RVBS phase can be prepared adiabatically. Our current protocol, which  is not optimized, requires a running time of       
 $10 J^{-1}$ while we estimate that coherence time can be 10 times larger. 
 For instance, in the setup of \cite{Bernier17} by setting $|\eta|=0.38 a_x= 1\mu$m  we get $J=C_6/a_x^6\sim 20$ MHz, such that the overall coherence time of the Rydberg system (estimated in \cite{Bernier17} to be $7 \mu$\ac{s})  
 is of the order of $100 J^{-1}$.  
 
A similar adiabatic protocol applies to the preparation of the RVBS phase in a $N_x \times N_y$ lattice. We fix the displacement of the atoms in the pairs to be $\eta=a_x(0.33\hat x + 0.38\hat z)$, 
and the displacement of the sublattices along $y$ to $d_y= 0.07a_x$ such that the average distance of the pairs along $y$ is $a_y= 0.88a_x$.   
With such values, the (blue-detuned) driven array is described by the Hamiltonian \eqref{eq:Heffspinsquare} with $\Lambda=-0.088 C_6/a^6_x$ and $G=1.42 C_6/a^6_x$. 
Since we get a similar ratio $\Lambda/G$ as in the periodic ladder, we expect the adiabatic preparation to work in a similar way.

\section{Conclusions and Outlook}\label{sec:outlook}
In this paper, we have shown that we can perform scalable quantum simulation of 2D lattice gauge theories with reconfigurable Rydberg arrays in current experiments.
As a prototype of gauge theory with magnetic (plaquette) interactions, we have targeted the RK model, 
a spin $1/2$ U(1) gauge theory that it is relevant for quantum magnetism. 
With the help of the electro-magnetic duality, we have evidenced that the dynamics of physical states has a blockade character that it is realized by geometrically tuned Rydberg arrays.
We have detailed the engineering of the dual RK Hamiltonian on the square lattice without background charges and we have computed its phase diagram for varying Rabi couplings. 
We have shown how to prepare and detect the RVBS phase in Rydberg experiments with ladders and 2D arrays.

Our findings open several new possibilities for the quantum simulation of lattice gauge theories and more generally of many-body physics.

\subsection*{Generalized RK Hamiltonians} 
The relation we have established between the RK model and driven Rydberg arrays through the duality extends also to other (i) lattice geometries and (ii) charge backgrounds \cite{Celi19}. 
Furthermore, the Rydberg implementation may suggest (iii) new mechanisms to achieve U(1)-spin liquid phases in two dimensions.

{\it Geometries.} Contrary to the quantum dimer counterpart \cite{Moessner11}, the phase diagram of the RK model without background charges has not received much attention on other lattice geometries than the square ones. 
The precise structure of the phase diagram is generally unknown and we may expect surprises in the phase diagram especially for frustrated lattices that are not bipartite. 
The first case to examine is RK Hamiltonian on a triangular lattice as its dual is realized by a properly deformed hexagonal decorated Rydberg array \cite{Celi19}.
 
{\it Charges.} Static background charges are especially interesting because they allow to probe confinement in the gauge theory and can lead to nested phases \cite[see Fig. 6]{Banerjee13}. 
In the dual RK model static background charges can be incorporated by modifying the map \eqref{eq:formalmap} and the blockade condition governing the plaquette flip \eqref{eq:map2D}. 
For instance, if we place a pair of two $\pm 1$ charges at distance we must include the effect of the string between the charges, which amounts to flip the link spins along the path 
(for very recent study on a special background see \cite{Herzog_Arbeitman19}). 
Thus, the map \eqref{eq:formalmap} acquires a minus sign along the path of string that changes the projector in \eqref{eq:map2D} along the string. 
Notice that by applying the same reasoning to the staggered distribution of charges that leads to the quantum dimer model (see appendix \ref{sec:qdimer}) 
one recovers the duality to the fully frustrated Ising model first found in \cite{Moessner00}.\\
As a final remark we notice that the dual approach considered here can be extended to simulate the Higgs mechanism in the untruncated U(1) gauge theory considering larger plaquette spins (see below) 
as in \cite{Zhang18}, and, perhaps, an analogue version of it for fermionic matter, cf. \cite{Zohar19}. 
\ac{In principle, the dual approach we consider here can be extended from static (c-number) to dynamical charges.
The price to pay is that the tunneling of charges become non-local in the charges, as in the 1D Schwinger model, 
and in the plaquette spins. The implementation of such a term would be possible in a digital, i.e. a Trotterized approach of the time evolution \cite{Weimer10}. 
Whether such term can be conveniently engineered within present Rydberg technology is under investigation.
Alternative digital schemes that include dynamical matter can be found in \cite{Zohar17,Bender18}.}

{\it U(1)-spin liquids.} It is well known that the Coulomb phase of compact U(1) gauge theory is stable in three dimensions while it is unstable in two dimensions due to instanton effect \cite{Polyakov87}. 
This continuum argument explains a posteriori why, contrary to 3D \cite{Moessner03,Hermele04}, the RK point of the RK model in 2D does not extend to a spin liquid phase on bipartite lattices, 
which is instead replaced by a RVBS phase that is confined. 
The absence of a deconfined ``photon'' excitation that rules out the existence of a spin liquid phase can be circumvented by breaking translational invariance 
as in happens with Cantor deconfinement \cite{Fradkin04,Vishwanath04,Papanikolaou07}. 
It is very interesting to explore whether in a similar spirit we can break translational invariance 
such to induce and stabilize a spin liquid phase in 2D, e.g. by considering additional interaction terms to the dual RK model that are natural from the Rydberg perspective. 
Alternatively, it is intriguing to explore the realization of a 3D spin liquid phase or of magnetic monopoles \cite{Castelnovo08} in 3D Rydberg arrays.

\subsection*{New experimental probes} 
Our approach to quantum simulations of the dual RK model allows i) to access experimentally quantum correlations and ii) to probe excitations' spectrum and thermalization of the RK model through quantum quenches, beyond
condensed matter experiments and classical computations. 

{\it Quantum correlations.} Contrary to traditional measurement schemes in condensed matter, Rydberg simulators allow for single-site resolution to detect whether each atom is in the ground or excited state.
On the one hand, such possibility allows us to characterize the phase diagram for different value of $\lambda$ through the structure factor and the expectation value of the generalized RK potential, which
are experimentally accessible observables. On the other hand, it allows us to access non-local order parameters like the Wilson loop or to measure directly the entanglement entropy by quantum interference as in \cite{Islam15}, 
or by random measurements as in \cite{Brydges19}, (for the theory proposal see \cite{Moura04,Daley12,Abanin12} and \cite{Elben18}, respectively). 
An alternative promising route to characterize quantum many-body states is to use quantum machine learning techniques, e.g for performing quantum many-body state reconstruction efficiently as proposed in \cite{Torlai18} 
and experimentally realized in \cite{Torlai19}, or to identify phases with an unknown, non-local order parameter \cite{Cong18}. 

{\it Dynamical probes.} Static quantum correlations discussed above can be potentially determined numerically for instance by quantum Monte Carlo or by DMRG 
for sufficiently large lattices such to approach the thermodynamic limit, see \cite{Banerjee13} and \cite{Tschirsich18} for recent calculations. 
A key advantage of the quantum simulators is that is equally easy to study the time evolution of the observables after a local or a global quench \cite{Keesling19}.
These experiments could shed light both on the excitations above the ground state and the thermalization properties in the gauge theory. In particular, it would be very interesting to study whether the relation between
confinement and many-body quantum scars \cite{Bernier17,Turner18,Ho19}, first noted for Rydberg chains and the 1D Schwinger model \cite{Surace19}, 
extends to 2D pure gauge theories (for dynamical phase transition in the Schwinger model see \cite{Zache19}). 
Recent numerical studies with exact diagonalization in 2D spin-1/2 U(1) gauge theory without charges in \cite{Huang19} and in the quantum dimer model and \cite{Lan18} have found evidences on small lattices 
of dynamical phase transitions (the emergence of kink-like structures  in the return amplitude
to the original ground state manifold)  for quantum quenches from flippable to RVBS phase and the emergence of glassy behavior (very slow relaxation time) for quantum quenches 
from flippable to the unflippable phase, respectively.
The latter behavior have been very recently confirmed and related to the constraint dynamics of the dimers in \cite{Feldmeier19}.

\subsection*{Exotic spin models} 
The use of configurable Rydberg arrays of atom pairs induces effective interacting spin-1/2 Hamiltonians with tunable couplings between spins. 
This is precisely the technique we have employed in the quantum simulation of the RK model. 
Following a similar strategy and clustering together $2S +1$ atoms in a macro-atom 
we can form composite spins $S$ whose spin-spin interactions are still controlled by the arrangements of the macro-atoms and their relative displacements. 
Therefore, configurable Rydberg arrays can be used for programmable simulations of exotic spin-$S$ models with anisotropic interactions.
We expect the local competition between ferromagnetic and antiferromagnetic interactions to lead to novel entangled phases and behavior. Additionally, the generalized blockade conditions that emerge in the simulator can encode 
known as well as novel ``quantum'' cellular automata 
\cite{fnote4}. 
The complexity generated by arranging driven Rydberg arrays in clusters can be viewed as the dynamical counterpart of quantum optimization for Maximum Independent set studied in \cite{Pichler18}.
We foresee that the combination of clustering of the arrays with angular dependent Rydberg-Rydberg interactions and/or with more complex Rydberg series such as those associated with earth-alkali atoms 
\cite{Cooper18,Norcia18,Saskin18} will open even more exciting perspectives for quantum simulation.

\acknowledgements{
AC thanks D. Gonzalez Cuadra for deriving the dual form of the plaquette operator of the spin-1/2 gauge theory on an open ladder 
and P. Silvi for discussions on the formulation of the same duality on square lattice. 
We thank A. Keesling and W.W. Ho for useful discussions and M. Dalmonte for insightful comments on an advanced version of the manuscript. 
Research in Innsbruck is supported by the ERC Synergy Grant UQUAM (grant agreement number 741541), the QuantEra project QTFLAG, the project PASQUANS of the EU Quantum Technology flagship, 
and the Simons foundation via the Simons collaboration UQM.
Research in Harvard is supported by the Center for Ultracold Atoms, the National Science
Foundation, the US Department of Energy and the Office of Naval Research.
AC acknowledges support from the UAB Talent Research program and from the Spanish Ministry of Economy and Competitiveness under Contract No. FIS2017-86530-P.
OV acknowledges support from RCC Harvard and Fundacion Ramon Areces. 

}

\appendix

\section{Spin gauge theory as dimer model}\label{sec:qdimer}

\begin{figure}[htb]
\begin{center}
\resizebox{0.9\columnwidth}{!}{\includegraphics{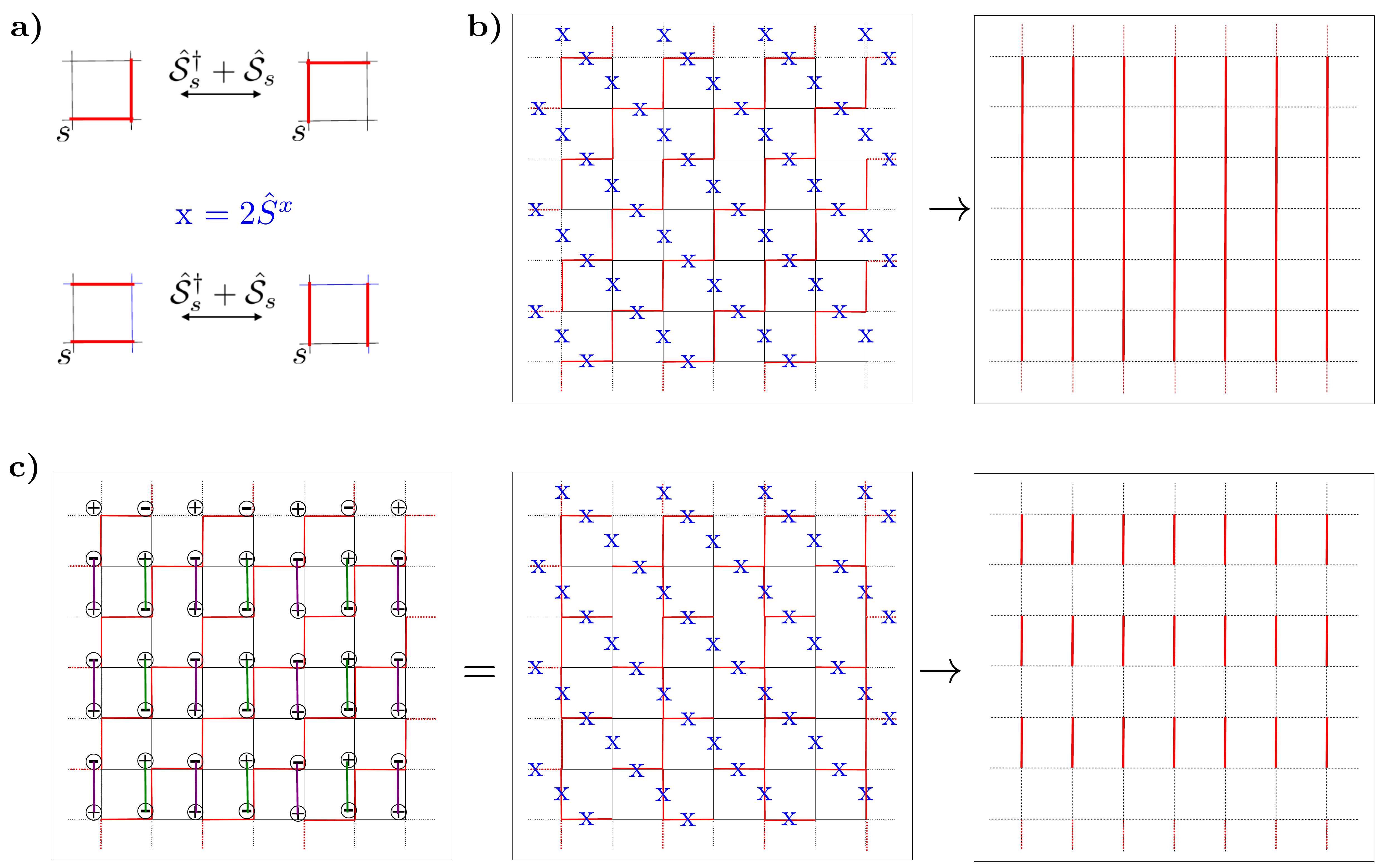}}
\caption{(color online) {\bf Relation between gauge magnets and quantum dimer model.} {\bf a)} By reversing the spins of the links at the bottom-left (or at the top-right) corner of the plaquette, 
the plaquette operator becomes a dimer move, see \eqref{eq:pldimer}.  {\bf b)} In the dimer basis, obtained by reversing the spins of the links at the bottom-left corner of the odd plaquettes, 
the fully flippable background $\ket{\Omega}$ appears as a columnar state. {\bf c)} The physical states of the quantum dimer model are obtained for a staggered distribution of static charges, $Q_s=(-1)^s$.
Such states can be constructed by applying the plaquette operator on the maximally flippable background depicted in the figure. It is obtained by superimposing a string covering associated to the charges on
$\ket{\Omega}$. Due to the charges, only half of the plaquettes are flippable.    
} 
\label{fig:qdimer}
\end{center}
\end{figure}

In this section we detail the relation between the spin-1/2 U(1) gauge theory and the quantum dimer model on a square lattice.
First of all, we notice that by applying a $\hat S^x$ transformation on the links at the bottom-left corner, $(s,x)$ and $(s,y)$, or on the links at the top-right one, $(s+\hat x,y)$ and $(s+\hat y,x)$ the 
plaquette operator on the plaquette $s$ assumes a ring exchange form with alternated spin flips, 
\begin{equation}
\hat{\cal S}_s^\dagger + \hat{\cal S}_s = \hat S^+_{s,x} \hat S^-_{s+\hat x,y} \hat S^+_{s+\hat y,x} \hat S^-_{s,y} + H.c.. \label{eq:pldimer}
\end{equation}
As a consequence the flippable plaquette configurations become the ones with alternated spins, i.e. the ones with parallel colored links, see Fig. \ref{fig:qdimer}a.
By extending the transformation to the full 2D plane, for instance, by reverting the bottom-left corner of the odd plaquettes (= the top-right corner of the even ones),
as shown in Fig. \ref{fig:qdimer}b, the RK Hamiltonian is still of the form \eqref{eq:HRK}, with the plaquette operator given in \eqref{eq:pldimer}, 
while the Gauss law \eqref{eq:GaussKS} takes becomes
\begin{equation}
\label{eq:GLspinice}
 \hat E_{s,x}+\hat E_{s,y}+\hat E_{s-\hat x,x} +\hat E_{s-\hat y,y} + (-1)^s \hat Q_s =0,
 \end{equation}
where $(-1)^s \equiv -1^{m+n}$, where $m,n$ are the cartesian coordinates of the site $s=m \hat x + n \hat y$. 
In Fig. \ref{fig:qdimer}b, we show that the background $\ket{\Omega}$ becomes a columnar state, the fully flippable background in the new basis.

The quantum dimer model is obtained by considering the physical states with staggered background charges.
For instance, by taking $Q_s=(-1)^s$ to each site is attached only one dimer that is identified by the colored link in $\ket \uparrow$, while the empty ones are in $\ket \downarrow$.
In Fig. \ref{fig:qdimer}c, we represent one of the maximally flippable configurations of the quantum dimer model: only half of the plaquettes are flippable. It can be visualized as the 
result of a special string covering superimposed over $\ket{\Omega}$. The string connecting the charges block at least half of the plaquette and change the physical states.
Several properties distinguish the physical states of the quantum dimer models from the physical states without background charges. Although they share 
the same Hamiltonian \eqref{eq:pldimer}, the different boundary conditions determine a different phase diagram and a different dual Hamiltonian for the two models.

\section{Electric truncation as a dynamical process and relation with Kogut-Susskind U(1)}\label{sec:truncation}
\ac{
We analyze here the dynamics of spin gauge theories in comparison to the  non-truncated U(1) gauge theory. For simplicity, we explicitly discuss the pure gauge case but the analysis applies also in presence 
of dynamical matter. We start by rewriting the pure gauge Kogut-Susskind Hamiltonian  \eqref{eq:HKS} in the rotating frame of the electric term
\begin{equation}
H'_{KS}(t) =  
\frac {-1}{2g^2}\sum_s \hat  V_{s,x}^\dagger(t) \hat V_{s+\hat x,y}^\dagger(t) \hat V_{s+\hat y,x}(t) \hat V_{s,y}(t) +H.c., \label{eq:HKSrot}
\end{equation}
where $V_{s,\mu}^{(\dagger)}(t) = \exp[ \tfrac {-i g^2}2 \hat E^2_{s,\mu}] \hat U_{s,\mu}^{(\dagger)} \exp[ \tfrac {i g^2}2 \hat E^2_{s,\mu}]$, $\mu=x,y$.
Since $\hat U_{s,\mu}^{(\dagger)}$ decreases (increases) the electric field by one, the magnetic couplings between highly excited electric states are fast rotating.  
Thus, in the rotating frame the usual competition between the electric and magnetic translates in a {\it soft} constraint for the plaquette operator that dynamically suppresses the coupling
between highly excited electric states.} 

\ac{
We can obtain the truncated theory dynamically from \eqref{eq:HKSrot} by deforming the electric term. For instance, we can achieve the spin $\frac 12$ link model Hamiltonian by first shifting the 
electric eigenvalues by $\frac 12$ (which is equivalent to include a $\theta$ term equal to $\pi$ in the U(1) Hamiltonian) and then considering a rescaled  electric term  
$\frac{g^2}2 \kappa \sum_s \left(\hat E_{s,x}^2+\hat E_{s,y}^2 -\frac 12\right)$. In the limit of $\kappa \to \infty$, the electric states with $|E|>\frac 12$ decouple and the plaquette operator reduces to \eqref{eq:plaquette}.
Such modification is equivalent to replace the quadratic electric term with one of a box potential form.  
Although the electric term acts as the identity on the remaining electric state, its effect on the dynamics persists: the plaquette operators on neighboring plaquettes do not compute due to the truncation of the electric states. 
The electric truncation in spin gauge theories is equivalent to a hard version of the energy penalty due to the electric interactions. 
The competition between electric and magnetic interactions that characterizes gauge theories in more than one dimension survives in the truncated theories. 
The spin $\frac 12$ link models we study here are, thus, the minimal instance of such gauge theories with continuous gauge groups.}

\section{Construction of the duality}\label{sec:formal}

Without charges, all gauge invariant electric states of the U(1) Kogut-Susskind lattice gauge theory can be written as combinations of closed electric loops, 
i.e. closed oriented paths of links with constant electric field (with respect to the path orientation). With the exclusion of ``large'' topologically non-trivial loops, every other loop configuration 
and thus all physical states can be written as combination of elementary plaquette loops. The electric field of a loop on the plaquette $p$ (conventionally oriented 
anticlockwise) assumes the meaning of a potential and it is called height, $h_p$. In two dimensions, the electric field on each link is given by the difference of the heights of the two  
plaquettes sharing that link (accordingly to the standard lattice orientation, it is the difference between the left and right heights). Therefore the electric field operator can be written as an operator relation
\begin{align}
\hat E_{p,x}&\to \hat h_p-\hat h_{p-\hat y}\cr
\hat E_{p,y}&\to   \hat h_{p-\hat x}-\hat h_p, \label{eq:heightmap}
\end{align}
which defines the dual theory on the plaquette basis.  
The heights can take any integer value and there is a unique identification between the heights and the electric states on the links, up to the definition of the overall height origin. 
The Kogut-Susskind Hamiltonian \eqref{eq:HKS} in the dual basis simplifies considerably as the plaquette operator, 
$\hat {\cal U}_p + \hat {\cal U}_p^\dagger = \hat U_{p,x} \hat U_{p+\hat x,y} \hat U_{p+\hat y,x}^\dagger \hat U_{p,y}^\dagger +H.c.$, 
acts on the plaquette $p$ 
by raising and lowering the height $h_p$ by one, 
\begin{equation}
[\hat h_p,\hat {\cal U}_p^{(\dagger)}]=-(+) \hat {\cal U}_p^{(\dagger)}.\label{eq:dualKSplaquette}
\end{equation} 

The height construction holds also in a truncated theory like the spin-1/2 gauge theory described in Sec. \ref{sec:GM}, with the only difference that the truncation limits also the admissible height values.
Since the value of the electric field on the links and thus the difference between the neighboring heights is limited to $\pm \frac 12$, the heights are identified mod. 2 
and can be represented as components of a spin 1/2. Thus, we can replace \eqref{eq:heightmap} with \eqref{eq:formalmap} given in Sec. \ref{sec:duality}.

Any allowed raising or lowering of the height is identified with a flip of the dual spin, thus from \eqref{eq:dualKSplaquette} we have  
$\hat{\cal U}_p,\hat{\cal U}_p^\dagger \to 2 \hat S^x_p$.
Furthermore, we can write the plaquette operator in the truncated theory as the Kogut-Susskind one projected on the allowed electric states
\begin{multline}
{\cal S}_p + {\cal S}_p^\dagger =  2 S^x_p \big(\frac 12 - \hat S^z_{p,x}\big)    \big(\frac 12 - \hat S^z_{p+\hat x,y}\big)   \\
                      \big(\frac 12 + \hat S^z_{p+\hat y,x}\big)\big(\frac 12 + \hat S^z_{p,y}\big) + H.c. . \label{eq:truncated_plaquette}
\end{multline}  
By expressing the projectors on the links in terms of the plaquette spins through \eqref{eq:formalmap} we find \eqref{eq:map2D}.

\section{Details on the Rydberg implementation}\label{sec:impledet}

Here we detail the implementation of the blockade condition on the square lattice.
For the array geometry illustrated in the left panel of Fig. \ref{fig:imple}a, we 
have by construction that $\sum_{p'=\langle p\rangle_1} B(p-p',\eta)=0 $. Indeed, from \eqref{eq:Bpeta} it follows that 
$B((a_y \pm d_y)\hat y,\eta)=0$  because the displacement is in $xz$-plane and $\eta \cdot \hat y =0 $, and $B(-\hat x\pm d_y \hat y,\eta) =- B(\hat x\pm d_y \hat y,\eta)$.    
The Ising couplings to the left and right NN spins are the same, $A(-\hat x \pm d_y \hat y,\eta)=A(+\hat x \pm d_y  \hat y,\eta)$,
while ones to the top and to the bottom NN spins are different,   $A((-a_y \pm d_y) \hat y,\eta)\neq A((-a_y \pm d_y) \hat y,\eta)$, for $d_y\neq 0$.
Thus, we can achieve the desired generalized blockade condition by solving    
\begin{equation}
-2A(+\hat x + d_y  \hat y,\eta) = A((a_y - d_y) \hat y,\eta) +A((a_y + d_y) \hat y,\eta),
\end{equation}
for $a_y=a_y(\eta,d_y)$. The solution exists for a wide range of the parameters $|\eta|$, $\theta$, and $d_y$.  
The function $a_y(\eta,d_y)$ cannot be written in closed form and has to be calculated 
numerically. Through $a_y(\eta,d_y)$ we can calculate both the gap  
\begin{multline}
G(\eta,d_y) = \text{Min}[A((a_y(\eta,d_y) + |d_y|) \hat y,\eta),\\
\frac 12(A((a_y(\eta,d_y) - |d_y|) \hat y,\eta)-A((a_y(\eta,d_y) + |d_y|) \hat y,\eta))],
\end{multline}
which determines the regime of validity of the effective RK Hamiltonian  \eqref{eq:Heffspinsquare},
and the coefficient of the modified RK potential \eqref{eq:V2square}
\begin{equation}
\Lambda(\eta,d_y)= A(\hat x + a_y(\eta,d_y), \eta).
\end{equation}

The best-suited values of the free parameters  $|\eta|$, $\theta$, and $d_y$ are obtained through the requirements that i) the spin effective description of the atom pair is valid, ii) the gap is maximal, 
such that the overall energy scale of the effective Hamiltonian is as higher as possible compared to the inverse decoherence time, and iii) the ratio $\Lambda/G$ is 
sufficiently small such that we can access the RVBS phase within the validity of the effective Hamiltonian \eqref{eq:Heffspinsquare}.
Since the intra-pair and outer-pair separation of the atoms strongly depends on the modulus of the displacement $|\eta|$, the first requirement sets an upper bound on $|\eta|$. 
This bound depends weakly on $\theta$ and $d_y$, and it is an increasing function of the former and decreasing function of the latter.
Since the gap $G$ strongly depends on $d_y$, the second requirement fixes an optimal value for $d_y$ that is weakly increasing function $|\eta|$ and $\theta$.
The value of $\Lambda$ strongly depends on $\theta$: in fact as it happens for dipolar interactions it exists a magic angle for which it is exactly zero. 
However, also the gap $G$ depends $\theta$ in a similar way: in this case the magic angle is smaller. In other words, we can diminish the ratio $\Lambda/G$ and satisfies the third requirement
by choosing sufficiently large $\theta$. An optimal compromise between a large gap $G$ and small $\Lambda/G$ ratio is obtained for $|\eta|=0.5a_x$, $d_y=0.07a_x$, and $\theta = 0.85 rad$, which gives $a_y=0.88a_x$,
$G=1.42 C_6/a_x^6$, and $\Lambda=-0.088 C_6/a_x^6$. Thus, at the price of reducing the overall energy scale (by an order of magnitude in relative terms) we can achieve a similar ratio $\Lambda/G<\frac 1{16}$ as in the periodic ladder 
and ensure the existence of an accessible RVBS phase. 
In absolute terms, by considering the setup of \cite{Bernier17} and by setting $|\eta|= 0.5 a_x = 1 \mu$m and $J=0.16 C_6/a_x^6\sim 16$ MHz, we estimate a coherence time (in \cite{Bernier17} of $7 \mu$\ac{s}) of      
the order of $80 J^{-1}$.   

\section{Details on the DMRG simulations}\label{sec:DMRG}
Here, we provide details on the DMRG simulations that were realized using the ITensor library. The 2D indices ($i_x,i_y$) entering the constructing of  Matrix-Product-Operators (MPO) were ordered as $l=i_y+(i_x-1)N_y$  
for $\mathrm{mod}(i_x,2)=1$, and $l=N_y+1-i_y+(i_x-1)N_y$  for $\mathrm{mod}(i_x,2)=0$. We consider periodic boundary conditions along the $y$-axis.

We used a small pinning field $\tilde \delta=0.1J$ along the $z$ direction, on three sites $(i_x,i_y)=(1,1),(2,1),(2,2)$  to favor one groundstate in the case of degeneracies (in particular for $\lambda\to-\infty$).
We also impose an energy penalty term to only select ground states within the physical subspace
\begin{align*}
H_\mathrm{pen} &= 
E \sum_{p} \left(P^\downarrow_p P^\downarrow_{p+x} P^\uparrow_{p+y} P^\uparrow_{p+x+y}
+P^\downarrow_p P^\uparrow_{p+x} P^\downarrow_{p+y} P^\uparrow_{p+x+y}
\right.\cr
&\left.+ 
 P^\uparrow_p P^\uparrow_{p+x} P^\downarrow_{p+y} P^\downarrow_{p+x+y}
+P^\uparrow_p P^\downarrow_{p+x} P^\uparrow_{p+y} P^\downarrow_{p+x+y}\right),
\end{align*}
with $P^{\uparrow,\downarrow}_p =(1/2 \pm S_p^z)$. One can check that the states that can be reached by the plaquette operators from the vacuum $\ket{\Omega}$, i.e., 
that satisfy the Gauss law in the dual formulation, satisfy $\langle H_\mathrm{pen} \rangle=0$.
In our simulations, we used $E=5J$ and $E=10J$.

Finally, we achieved ground state convergence,  below the percent level w.r.t spin-spin correlations,  for maximum bond dimensions $D=256$.

\begin{figure}[t]
\resizebox{0.95\columnwidth}{!}{\includegraphics{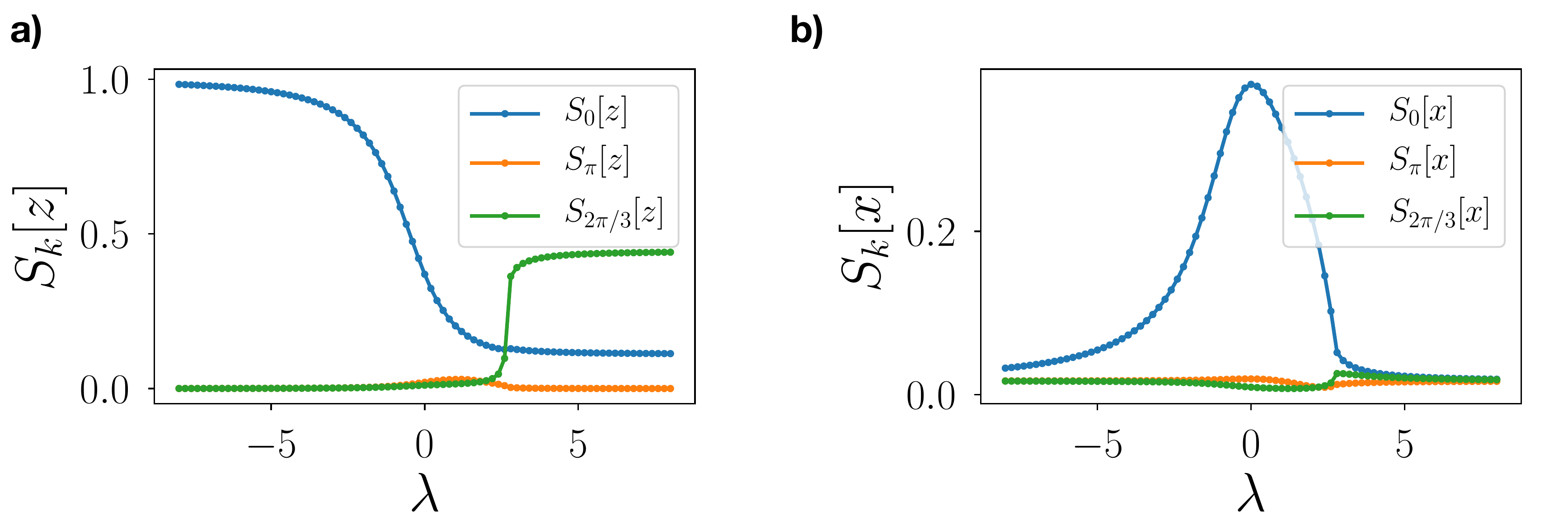}}
\caption{(color online) {\bf The ground state of the 1D PXP model.} 
{\bf (a)} Structure factors for $z$ spin-spin correlations for $k=0,\pi,2\pi/3$. 
{\bf (b)} Structure factors for $x$ spin-spin correlations for $k=0,\pi,2\pi/3$.
Parameters: $N_x=60$, maximum bond dimension $D=128$. 
} 
\label{fig:1D}
\end{figure}

\section{Ground state of the PXP model}\label{sec:PXP}
In this section, we show results for the ground state of the 1D PXP model
\begin{equation}
H_{1D}= \sum_p P_{p-\hat x}^\uparrow P_{p+\hat x}^\uparrow (-2S_p^x+\lambda),
\end{equation}
calculated via DMRG, with open boundary conditions, and with a small pinning field $\tilde \delta=0.1 J$ on the first site. 
We also impose the blockade constraint $ \sum_p \langle  P^\downarrow_p P^\downarrow_{p+x} \rangle$ by energy penalty. 

The phase diagram in terms of structure factors is represented in Fig.~\ref{fig:1D}, and shows no signatures of a RVBS phase.


\begin{thebibliography}{10}
\bibitem{notescalable} {\footnotesize{Recent experiments \cite{deMello19} show that one million atom arrays may become possible in the foreseeable future.}}
\bibitem{fnote1} {\footnotesize Charge conjugation implies that the electric field and thus the link spins are
reversed.}
\bibitem{fnote2} {\footnotesize Contrary to what occurs for the square lattice, such Hamiltonians cannot be simply mapped into U(1) gauge theories of the form \eqref{eq:HRK} on the corresponding lattices.}
\bibitem{fnote3} {\footnotesize The duality introduced here is different than the one first noticed in \cite{Moessner00} 
between the dimer models on the square and hexagonal lattices and fully frustrated Ising model on their dual lattices. 
In fact, the realization of the generalized blockade Hamiltonian in terms of Rydberg atoms of Sec. \ref{sec:imple} is equivalent to the realization of
an anisotropic Ising model with first and second neighbor couplings on properly tuned geometries.}
\end{thebibliography}
\end{document}